\documentclass[aps,reprint,twocolumn,superscriptaddress,10pt]{revtex4-2}
\usepackage{siunitx}
\usepackage{amsfonts}
\usepackage{amsmath}
\usepackage{amssymb}
\usepackage{bm}% bold math
\usepackage{graphicx}
\usepackage{dcolumn}
\usepackage{mciteplus}
\usepackage{euscript}
\usepackage{multirow}
\usepackage{color}
\usepackage{textcomp}
\usepackage{gensymb}
\usepackage{dblfloatfix}
\usepackage{subfiles}
\graphicspath{{./figs/},{../figs/}}

\DeclareRobustCommand*\cal{\@fontswitch\relax\mathcal}

\newcommand{\ef}{E_{\rm F}}

\def\bfS{{\bf S}}
\def\bfm{{\bf m}}
\def\bfk{{\bf k}}
\def\bfq{{\bf q}}

\def\bfR{{\bf R}}
\def\bfe{\hat{\bm e}}

\def\abinitio{\emph{ab initio}}
\def\a{\alpha}
\def\b{\beta}

\def\nn{\nonumber\\}

\def\d{\delta}

\newcommand{\srmnsb}{\text{SrMn}\text{Sb}_2}

\newcommand\jij{J_{ij}}

\newcommand{\ud}{\,\mathrm{d}}

\newcommand{\redbf}[1]{}

\newcommand{\memome}[1]{}
\newcommand{\req}[1]{Eq.~(\ref{#1})}

\newcommand{\rfig}[1]{Fig.~\ref{#1}}
\newcommand{\rFig}[1]{Figure~\ref{#1}}
\newcommand{\rtbl}[1]{Table~\ref{#1}}

\begin{document}

\title{Magnetic interactions and excitations in SrMnSb$_2$}

\author{Zhenhua Ning}
\affiliation{Ames National Laboratory, U.S. Department of Energy, Ames, Iowa 50011}
\author{Bing Li}
\affiliation{Ames National Laboratory, U.S. Department of Energy, Ames, Iowa 50011}
\author{Weilun Tang}
\affiliation{Ames National Laboratory, U.S. Department of Energy, Ames, Iowa 50011}
\author{Arnab Banerjee} 
\affiliation{Department of Physics and Astronomy, Purdue University, West Lafayette, IN 47906}
\author{Victor Fanelli} 
\author{Doug Abernathy} 
\affiliation{Neutron Scattering Division, Oak Ridge National Laboratory, Oak Ridge, TN, 37831}
\author{Yong Liu} 
\author{Benjamin G Ueland}
\affiliation{Ames National Laboratory, U.S. Department of Energy, Ames, Iowa 50011}
\author{Robert J. McQueeney}
\affiliation{Ames National Laboratory, U.S. Department of Energy, Ames, Iowa 50011}
\author{Liqin Ke}
\affiliation{Ames National Laboratory, U.S. Department of Energy, Ames, Iowa 50011}

\begin{abstract}
The magnetic interactions in the antiferromagnetic (AFM) Dirac semimetal candidate SrMnSb$_2$ are investigated using \textit{ab initio} linear response theory and inelastic neutron scattering (INS).
Our calculations reveal that the first two nearest in-plane couplings ($J_1$ and $J_2$) are both AFM in nature, indicating a significant degree of spin frustration, which aligns with experimental observations.
The orbital resolution of exchange interactions shows that $J_1$ and $J_2$ are dominated by direct and superexchange, respectively.
In a broader context, a rigid-band model suggests that electron doping fills the minority spin channel and results in a decrease in the AFM coupling strength for both $J_1$ and $J_2$.
To better compare with INS measurements, we calculate the spin wave spectra within a linear spin wave theory, utilizing the computed exchange parameters.
Although the calculated spin wave spectra somewhat overestimate the magnon bandwidth, they exhibit overall good agreement with measurements from INS experiments.
\end{abstract}

\date{\today}
\maketitle

\section{Introduction}

Transition-metal dichalcogenides have the potential for intertwined charge, structural, and magnetic states~\cite{Wilde2022prb}. 
In particular, the $A$Mn$X_2$ (112), $A=$~Ca, Sr, Ba, Eu, or Yb and $X =$~Sb or Bi, have generated intense contemporary interest because they exhibit perfect or slightly-distorted square magnetic Mn layers predicted to support Dirac or Weyl fermions~\cite{Hasan2010rmp,Bansil2016rmp,Armitage2018rmp,Lv2021rmp,Vanderbilt2018}. 
The coupling of magnetic order or fluctuations to such topological quasiparticles is compelling as magnetic control may allow for tuning of topological properties.  
Thus, understanding the basic magnetic interactions in the 112 compounds is important.

The Dirac semimetal candidate SrMnSb$_2$ features a slightly-distorted tetragonal structure and exhibits C-type antiferromagnetic (AFM) ordering, as shown in Figs.~\ref{fig:crystalstructure}(a) and \ref{fig:crystalstructure}(b). 
The C-type order consists of N\'{e}el-type AFM order within the slightly-distorted square Mn sublattices and ferromagnetic (FM) along the out-of-plane direction between them.
The magnetic interactions and excitations of this compound have been studied using a combination of inelastic neutron scattering (INS) and density functional theory (DFT)~\cite{zhang2019prb}.
The study considered the nearest-neighbor (NN) exchange coupling parameter $J_1$, obtained from total energy mapping of various magnetic configurations. 
However, due to the semi-metallic nature of SrMnSb$_2$, it is reasonable to expect that the magnetic coupling extends to further neighbors.  

To provide a more comprehensive picture, more recent work on various systems has expanded to the minimalistic Heisenberg $J_1$-$J_2$ model~\cite{ceccatto1992prb, sirker2006prb}, which factors in the next-nearest-neighbor (NNN) coupling $J_2$. This model offers an elegant and straightforward framework for interpreting various spin configurations resulting from the competition between $J_1$ and $J_2$, taking into account their relative amplitudes and signs.
Experimentally, valuable information about the values of $SJ$ and the ratio of $J_1$/$J_2$ can be obtained through INS. However, further theoretical investigation is desirable, as it can offer a microscopic understanding of these interactions and guide the manipulation of $J_{ij}$, where $J_{ij}$ is the parameter corresponding to the exchange coupling between sites $i$ and $j$, to either alleviate spin frustration and tune magnetic order, or promote spin frustration and quantum spin fluctuations that can mediate electronic pairing.

\textit{Ab initio} estimations of exchange parameters can be performed using different methods, including total energy mapping and linear response theory~\cite{liechtenstein1987jmmm, kotani2008jpcm}.
The total-energy-mapping method is widely employed due to its simplicity.
For isotropic exchanges, the total energies of various collinear spin configurations are often calculated in first-principles methods and mapped onto a model spin Hamiltonian.
Interactions beyond the isotropic exchanges, such as the off-diagonal part or the anisotropic diagonal part of the $3\times3$ exchange tensor $J$, can become important for more complicated materials systems that have large spin-orbit coupling (SOC) and/or broken inversion symmetry~\cite{ye2021prb}.
To estimate the relativistic-effect-originated exchanges, such as anisotropic exchange (anisotropic diagonal part of $J$ tensor) or the Dzyaloshinskii-Moriya interaction (antisymmetric off-diagonal part of $J$ tensor), SOC needs to be included in \textit{ab initio} calculations, and non-collinear spin configurations need to be considered.
However, the mapping method has limitations.
For example, extracted interaction parameters can be non-unique due to the dependence on assumed model Hamiltonians.
Moreover, its applicability is limited in itinerant systems when the variability of on-site spin moments across different configurations, particularly in metallic compounds, becomes significant.

On the other hand, linear response theory evaluates the energy variations resulting from infinitesimal spin rotations away from the ground state.
It is more computationally demanding but, in principle, more suitable for the calculation of low-temperature spin-wave (SW) excitations, which can be regarded as a small perturbation to the ground state.
The linear response method is also easily extensible to beyond-DFT methods that may be challenging to access the total energy, such as many-body-perturbation-theory-based $GW$ methods~\cite{van-schilfgaarde2006prl}, to better describe electronic structures and magnetic interactions in various materials~\cite{kotani2008jpcm, ke2011prbr, ke2021ncm}.
Finally, the linear response method allows for easy resolution of pairwise interactions into orbitals and band-filling effects, revealing the microscopic origin of these exchange couplings and providing guidance for bandstructure engineering using doping and pressure.

Here, we apply linear response theory to investigate the magnetic interactions and excitations in $\srmnsb$.
We calculate and resolve isotropic exchange couplings, gaining insight into the microscopic origin of the magnetism.
Using the obtained exchange parameters, we calculate the SW spectra via linear spin wave theory (LSWT) and compare them with INS measurements.
This comparison confirms the substantial magnetic frustration existing in the Mn layers.
The linear-response results allow a discussion of how to tune the magnetic frustration by carrier doping.

\section{Computational and experimental methods}
We first construct the real-space scalar-relativistic tight-binding (TB) Hamiltonian $H(\mathbf{R})$ using the maximally localized Wannier functions (MLWFs) method~\cite{marzari1997prb,souza2001prb,marzari2012rmp}.
The reciprocal-space Hamiltonian $H(\mathbf{k})$ is obtained through Fourier transform.
To better address the potential influences from the slightly-distorted tetragonal structure, we symmetrize the TB Hamiltonian to ensure that the orthorhombic crystal symmetry is rigorously satisfied.
Afterward, the corresponding Green's function $G(\mathbf{k},\omega)$ is constructed for use in the linear response approach, as implemented in our recently developed TB Green's function code~\cite{ke2019prb}, to calculate the exchange couplings.
Finally, with the exchange parameters in hand, we proceed to construct SW spectra using a LSWT.

\subsection{Crystal structure}

\begin{figure}[h]
\centering
\begin{tabular}{c}
\includegraphics[width=1.0\linewidth,clip]{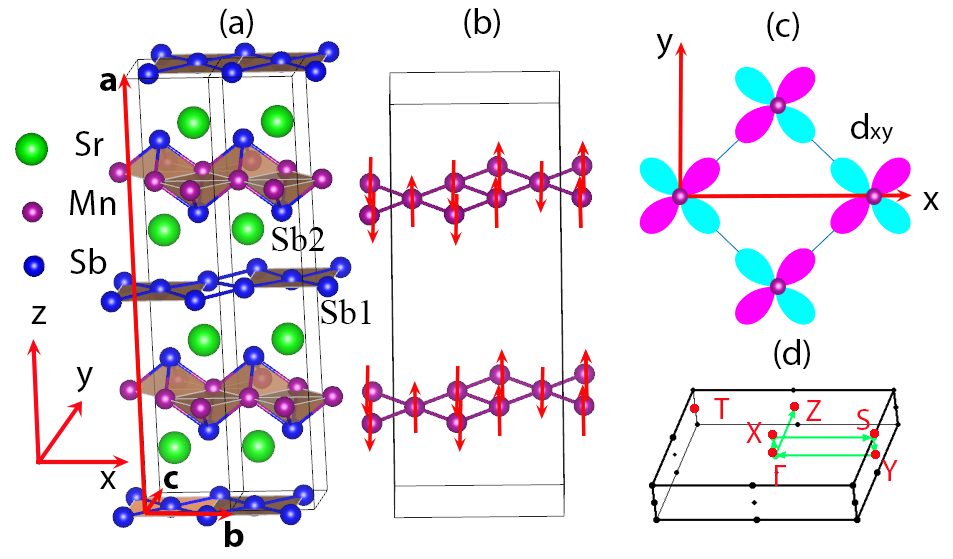} \\
\end{tabular}%
\caption{
  Crystal structure and corresponding Brillouin zone of $\srmnsb$.
(a) Schematic representation of the crystal structure of $\srmnsb$.
The primitive unit cell is doubled along the $\bold{b}$ direction to better illustrate the Mn grid.
Sr, Mn, and Sb atoms are represented by green, purple, and blue spheres, respectively.
The lattice vectors $\bold{a}$, $\bold{b}$, and $\bold{c}$ are highlighted in red.
For convenience in discussing the orbital contributions to exchange coupling, we align the longest lattice vector ${\bold a}$ along the $\hat{z}$ direction,
and the nearly-square Mn sublattice is on the ${\bold b}{\bold c}$-basal plane, or equivalently, the ${xy}$-plane.
(b) C-type AFM ordering of spin moments of Mn atoms in the cell shown in (a). 
(c) Top view of $d_{xy}$ orbitals for the Mn net shown in (b).
(d) The first Brillouin zone of $\srmnsb$ with high-symmetry $k$ points marked by red dots and $k$ paths marked by green arrows.
$X=[\frac12,0,0]$, $Y=[0,\frac12,0]$, and $Z=[0,0,\frac12]$ in orthorhombic reciprocal lattice units (r.l.u.).
}
\label{fig:crystalstructure}
\end{figure}

$\srmnsb$ crystallizes in the orthorhombic SrZnSb$_2$-type ($Pnma$, space group no.62) structure with lattice parameters $a = \SI{23.19}{\angstrom}$, $b = \SI{4.42}{\angstrom}$, and $c = \SI{4.46}{\angstrom}$~\cite{brechtel1981jlcm}.
The primitive cell contains four formula units (f.u.) and all atoms occupy different sets of $4c$ (.$m$.) sites.
The crystal structure is shown in \rfig{fig:crystalstructure}(a).
For convenience in discussing the Mn-$3d$ orbital contributions to the magnetic properties, we align the longest lattice vector $\mathbf{a}$ along the $\hat{z}$ direction, so the magnetic Mn layers are in the $xy$ ($\mathbf{bc}$) basal plane.
We also double the unit cell along the $\bold{b}$ direction to better illustrate the Mn grid.
The slight distortion of the $\mathbf{bc}$-basal plane results in the NNN exchange coupling $J_2$ becoming anisotropic with respect to the $\hat{x}$ ($\mathbf{b}$) and $\hat{y}$ ($\mathbf{c}$) directions, denoted as $J_{2b}$ and $J_{2c}$, respectively.

There are two Sb sublattices, denoted in \rfig{fig:crystalstructure}(a) as Sb$_1$ and Sb$_2$, respectively.
Two Sb$_2$ layers sandwich a Mn layer, forming MnSb$_4$ tetrahedra.
The Sr atoms are relatively weakly bonded to the Sb$_1$ atoms and staggered above and below the Sb$_1$ layer, forming rhombus nets~\cite{ramankutty2018sp}.
The C-type magnetic ordering lowers the crystal symmetry and separates the otherwise four equivalent Mn atoms into two sublattices, each possessing the same on-site magnetic moment but opposite orientation, as shown in \rfig{fig:crystalstructure}(b).

\rFig{fig:crystalstructure}(c) illustrates Mn-$d_{xy}$ orbitals pointing along the nearest neighbor direction, while \rfig{fig:crystalstructure}(d) shows the first Brillouin zone (BZ) with special $k$ points denoted.
Due to the large separation between Mn layers, the super-superexchange between Mn layers is much weaker than the intralayer coupling, similar to the magnetic topological insulator MnBi$_2$Te$_4$~\cite{li2021prb}.
Such quasi-2D magnetic structures generally require beyond-mean-field approaches to estimate critical temperatures~\cite{mkhitaryan2021prb}.
Experimental lattice constants and atomic position parameters~\cite{brechtel1981jlcm} are used for all calculations in this work.

\subsection{DFT details and TB Hamiltonian symmetrization}

DFT calculations are performed using the generalized-gradient approximation (GGA) with the Perdew-Burke-Ernzerhof exchange-correlation parametrization~\cite{perdew1996prl} and the projector augmented wave approach~\cite{kresse1999prb, kresse1996prb} as implemented in the Vienna \textit{ab initio} simulation package (\textsc{Vasp}).
To facilitate our subsequent analyses, we construct the MLWFs through a postprocessing procedure~\cite{marzari1997prb, souza2001prb, marzari2012rmp}, as implemented in \textsc{Wannier90}~\cite{mostofi2014cpc}, using the output of the self-consistent scalar-relativistic DFT calculation.
In total, the TB basis comprises 44 MLWFs, encompassing $3d$ orbitals for four Mn atoms and $5p$ orbitals for eight Sb atoms in the unit cell.
To minimize the spread functional for entangled energy bands, we adopt a two-step procedure~\cite{souza2001prb}.
For each spin channel, a real-space Hamiltonian $H^\sigma(\bf{R})$ with dimensions 44$\times$44 is constructed to accurately represent the band structures in a specified ``frozen'' energy window near $\ef$.
The energy bands are recalculated within TB to ensure that DFT bands can be accurately reproduced before further magnetic property calculations.

The process of Wannierization for DFT bands may not always preserve the symmetry of wavefunctions and orbital characteristics.
As we are also interested in understanding the potential impacts of the slightly-distorted square Mn grid on exchange couplings, we symmetrize the Hamiltonian accordingly using
\begin{equation}\label{hksymmed}
  H_{ij,\sigma}^\text{sym}(\bfR)=\dfrac{1}{|\mathcal{G}_H|}\sum_{\mathcal{R}\in \mathcal{G}_H} \langle \mathbf{0} |\hat{\mathcal{R}} \hat{H} \hat{\mathcal{R}}^{\dagger} | \mathbf{R} \rangle _{ij,\sigma}\,,  
\end{equation}
where $\mathbf{0}$ denotes the central unit cell, $\bold R$ denotes primitive translation vectors, $i$ and $j$ denote two sites within the primitive cell,  $\sigma$ is the spin channel, and $\mathcal{R}$ represents symmetry operations within the subgroup $\mathcal{G}_H$ of the Hamiltonian.
A detailed implementation of Hamiltonian symmetrization can be found in Appendix~\ref{symham}.

Before delving into magnetic properties calculations, we ensure alignment between the band structures generated by DFT and those produced by the symmetrized TB Hamiltonian:

\begin{equation} 
  H_{ij}^{\sigma}(\bold{k}) = \sum_{\bold R} H_{ij}^{\sigma}(\bold{R}) e^{-i \bold{k}\cdot \bold{R}}\,,
\label{hk}  
\end{equation}
where $\bold{k}$ denotes wave vectors.

\subsection{Exchange couplings, spin wave and critical temperature}

The static linear response method with the long-wave approximation~\cite{liechtenstein1987jmmm}, based on Green's function technique, has long been developed to calculate exchange couplings as defined in a Heisenberg-type Hamiltonian:
\begin{equation}
  H = -\sum_{i\ne j} J_{ij}^{e}\, \bfe_i \cdot \bfe_j.
\label{eq:h-je}  
\end{equation}
Here, $J_{ij}^e$ is the isotropic exchange interaction parameter between sites $i$ and $j$ in the crystal, and $\bfe_i$ is the unit vector pointing along the direction of the local spin moment at site $i$ in the reference spin configuration.
The early implementations utilized local-basis DFT methods, such as the linear muffin-tin orbital method~\cite{van-schilfgaarde1999jap}, employing the atomic sphere approximation (ASA) and based on Green's function technique.
These methods provided highly efficient and useful descriptions of magnetic interactions in various systems, particularly those with close-packed structures.
For example, they have been successfully applied to systems such as K$_2$Fe$_{4+x}$Se$_5$~\cite{ke2012prbr}, {${T}_{2}{\mathrm{AlB}}_{2}$ ($T$=Fe, Mn, Cr, Co, and Ni) and their alloys}~\cite{ke2017prb}, and $R$(Fe$_{1-x}$Co$_x$)$_{11}$Ti$Z$ ($R$=Y and Ce; $Z$=H, C, and N)~\cite{ke2016prb}.
Additionally, more accurate full-potential methods have also been implemented to evaluate $J_{ij}$ from the inverse static transverse susceptibilities, using the rigid-spin approximation that projects susceptibility onto the spin density of the magnetic sites~\cite{kotani2008jpcm,ke2013prb,ke2021ncm}.
Lastly, modern plane-wave-based DFT methods have often been interfaced with the MLWFs method to generate realistic TB Hamiltonians that accurately describe the band structures within an energy window typically near the Fermi level ($\ef$) of a material.
This approach also offers an attractive means to evaluate and analyze band structures and various other properties~\cite{pizzi2020jpcm, korotin2015prb, blanco-rey2019njp}, with both efficiency and accuracy.

We carry out the linear response calculations using our recently-developed TB code, which has been employed to efficiently analyze band structures~\cite{rosenberg2022prb,lee2023prb}, Fermi surface~\cite{timmons2020prr}, and magnetocrystalline anisotropy~\cite{ke2019prb}.
Starting from the TB Hamiltonian $H(\bfk)$, we construct intersite Green's function $G_{ij}^{\sigma}(\bfk)$ on a $16\times16\times4$ $k$ mesh to compute the exchange parameters $J_{ij}^e({\bf q})$.
Their orbital-resolved components $J_{ij,m}^e$ are defined by
\begin{eqnarray}
  J_{ij,m}^e(\bold{q}) &=& -\dfrac{1}{4\pi}\Im\int^{\ef}_{-\infty} \ud\omega \int_\text{FBZ} \ud\bfk \Big[ \\
    & & \Delta_{i}(\bfk)G_{ij}^\downarrow(\bfk,\omega)\Delta_{j}(\bfk+\bfq)G_{ji}^\uparrow(\bfk+\bfq,\omega)\Big]_{mm}\,, \nonumber
  \label{jij}  
\end{eqnarray}  
where the exchange-splitting matrix $\Delta_i(\bfk)=H_{ii}^{\uparrow}(\bfk)-H_{ii}^{\downarrow}(\bfk)$.
The real-space exchange constants $J_{ij}^e({\bf R})$ are then obtained through a subsequent Fourier transform.

It is important to note that the neutron scattering community often adopts a different convention, wherein the Heisenberg Hamiltonian is frequently defined in terms of the spin vector $\bfS$ as follows:
\begin{equation}
H=\sum_{i<j} J_{ij}^\text{N}\, \bfS_i \cdot \bfS_j.
\label{eq:h-jn}
\end{equation}
For clarity and to facilitate comparison between calculations and experiments, by comparing Eqs.~(\ref{eq:h-je}) and (\ref{eq:h-jn}), we have
\begin{equation}
  J_{ij}^\text{N}=-\frac{8}{\bfm_i \bfm_j}J_{ij}^{e}.
  \label{eq:jn-js}
\end{equation}
Here, $\bfm_i=2\bfS_i$ is the magnetic moment (with sign) on site $i$.
A positive (negative) $J_{ij}^\text{N}$ indicates AFM (FM) coupling.
In contrast to $J_{ij}^\text{N}$, $J_{ij}^e$ is defined with respect to the given spin configuration, and a positive (negative) $J_{ij}^e$ indicates that the given ordering of the corresponding pair is stable (frustrated).
The critical temperatures, Curie temperature for FM materials or N\'{e}el temperature for AFM materials, can be estimated within the mean-field approximation as $T_\text{C}=\frac{2}{3}J_0^e$, where $J_0^e=\sum_{i} J^{e}_{0i}$.
However, the mean-field approximation tends to overstate the critical temperature, especially in quasi-2D magnetic structures with very weak interlayer coupling.
All $J_{ij}$ discussed hereafter, unless specified otherwise, refer to $J_{ij}^\text{N}$.

With the $\abinitio$ exchange parameters obtained, the SW spectra can be calculated by solving the equation of motion method or through the bosonization of the spin Hamiltonian with Holstein-Primakoff~\cite{holstein1940pr} or other transformations that typically retain the two-boson terms in LSWT~\cite{toth2015jpcm}.
In the latter approach, higher-order four-boson terms can be included~\cite{mkhitaryan2021prb} to account for magnon-magnon interactions at finite temperatures.

\subsection{INS experiment and analysis}
The INS experiment was conducted on the wide angular-range chopper spectrometer (ARCS)~\cite{Abernathy2012} at the Spallation Neutron Source at Oak Ridge National Laboratory. 
For this experiment, 19 pieces of single crystals with a total mass of 386 mg were co-aligned. 
Using an optical microscope with polarized light, twin domains due to the slightly distorted orthorhombic unit cell, each rotated by 90$\degree$ with respect to the other, were observed on single-crystal pieces. 
Therefore single crystals were treated as pseudo-tetragonal while using our x-ray Laue camera to achieve co-alignment. 
The horizontal scattering plane was defined as $(H,K,L)$, where the momentum transfer $\mathbf{q}= H\mathbf{a}^*+ K\mathbf{b}^*+ L\mathbf{c}^*$ is in orthorhombic reciprocal lattice units (r.l.u.).
The rocking scans of the co-aligned assembly yielded full widths at half maximum of $3$ degrees. 
The measurements were performed with the incident neutron energy of $E_{\text{i}}=50$~meV or 125~meV at a temperature of $T=20$~K. 

To analyze the INS data, we performed least-squares fits and simulations utilizing the software package \textsc{pyLiSW}~\cite{pyLiSW}.

\section{Results and discussions}

We begin by calculating the total energies of $\srmnsb$ with various magnetic orderings, including FM as well as C-, G-, A-, and Stripe-type AFM configurations.
Our calculations confirm that among all the configurations, the C-type AFM ordering exhibits the lowest energy.

The MLWFs included in the TB Hamiltonian are determined by investigating the bandstructure characters near the $\ef$ in DFT.
According to DFT calculations, the Mn-$3d$ states dominate in the vicinity of $\ef$ and in the energy range of \SIrange{-4.0}{-2.5}{eV} in the majority spin channel and \SIrange{0}{1.5}{eV} in the minority spin channel, exhibiting a spin splitting of about \SI{3.8}{eV}.
The more dispersive Sb-$5p$ states also contribute to the bandstructure in the relevant energy window.
On the other hand, the unoccupied Sr-$4d$ states are located approximately \SI{3}{eV} above $\ef$ and can be neglected.
Consequently, the TB basis consisting of Mn-$3d$ and Sb-$5p$ Wannier orbitals provides a reasonable description of the electronic structure in the vicinity of $E_{\text{F}}$.

As noted above, the Wannierization procedure, in general, may not preserve the system's symmetry.
The basis Wannier functions might center at positions that deviate from the atomic centers, and their orbital characteristics may not be preserved.
We observe that the non-symmetrized TB Hamiltonian, based on MLWFs, slightly breaks symmetry of $\srmnsb$.
Therefore, we symmetrize the TB Hamiltonian using \req{hksymmed} before conducting magnetic property calculations.

\subsection{Electronic band structure and magnetization for $\srmnsb$}

\begin{figure}[bht]
\centering
\begin{tabular}{c}
\includegraphics[width=.80\linewidth,clip]{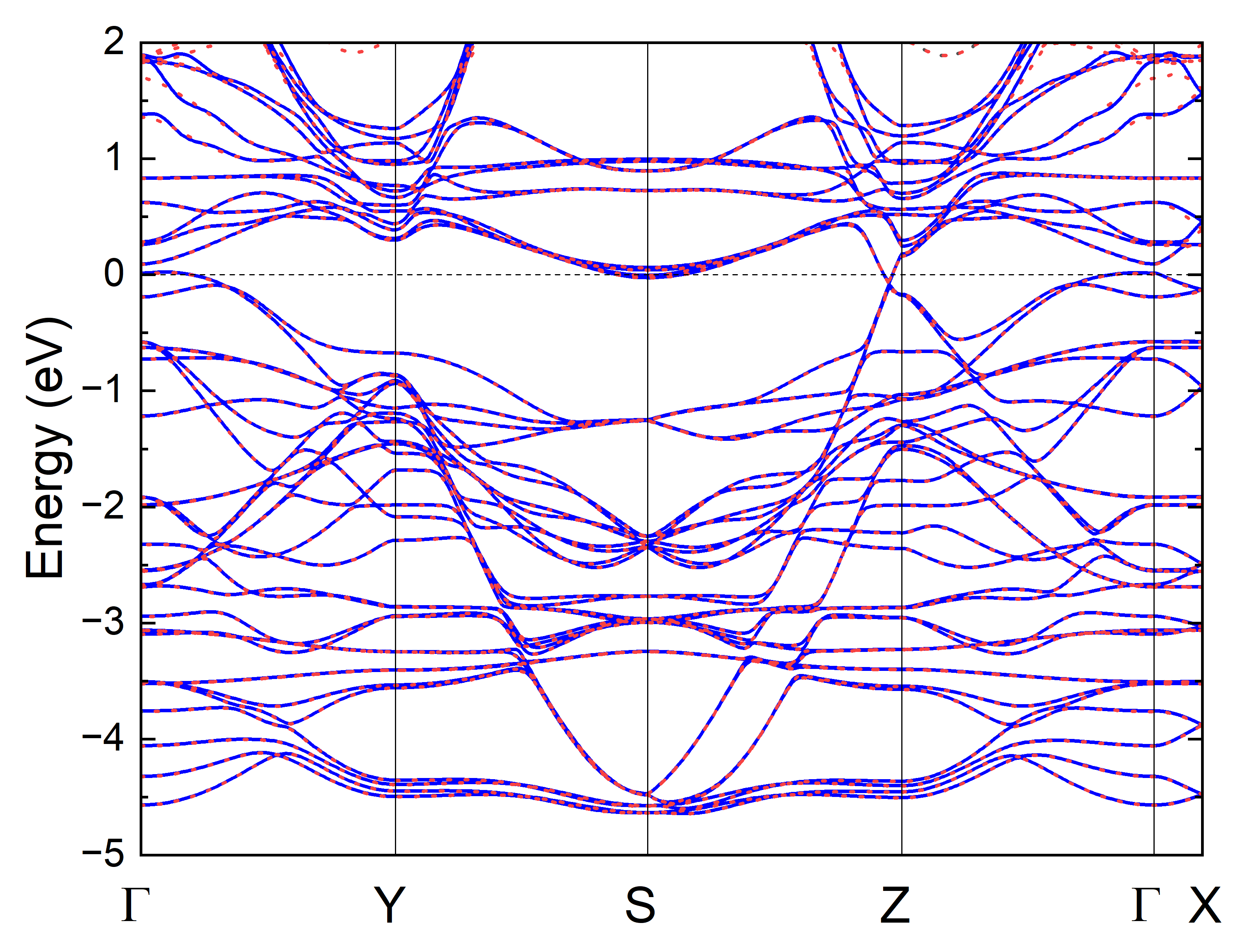}
\end{tabular}%
\caption{Band structure of $\text{SrMnSb}_2$ calculated using tight-binding (solid blue lines) and density functional theory (red dots).
  They agree perfectly below 1 eV within this energy window.
  The horizontal dashed line denotes the Fermi level.
}
\label{fig:band_tb_vs_dft}
\end{figure}

\rFig{fig:band_tb_vs_dft} compares the electronic band structures along the $k$-path $\Gamma$--Y--S--Z--$\Gamma$--X, calculated using the symmetrized TB Hamiltonian (in blue) and DFT (in red).
The excellent agreement in this energy window allows for a reasonable description of magnetic properties using the TB Hamiltonian.

With five electrons in the $3d$ shell, Mn atoms have a nearly fully-occupied majority spin channel and a slightly occupied minority spin channel, resulting in a large magnetic moment.
The calculated Mn on-site magnetic moment of \SI{3.81}{\mu_B/Mn} in DFT agrees well with the previous experimental value of \SI{3.78}{\mu_B/Mn}~\cite{liu2017nm,zhang2019prb}, and the DFT values of
\SIrange{3.75}{3.78}{\mu_B/Mn}~\cite{islam2020prb,zhang2019prb}.
Furthermore, the symmetrized TB Hamiltonian gives \SI{3.76}{\mu_B/Mn}, agreeing well with the DFT and experimental results.

\subsection{Magnetic interactions: Orbital contributions and Bandfilling effects}\label{sec:jijdiscus}

\begin{figure}[ht]
\centering
\begin{tabular}{c}
\includegraphics[width=.90\linewidth,clip]{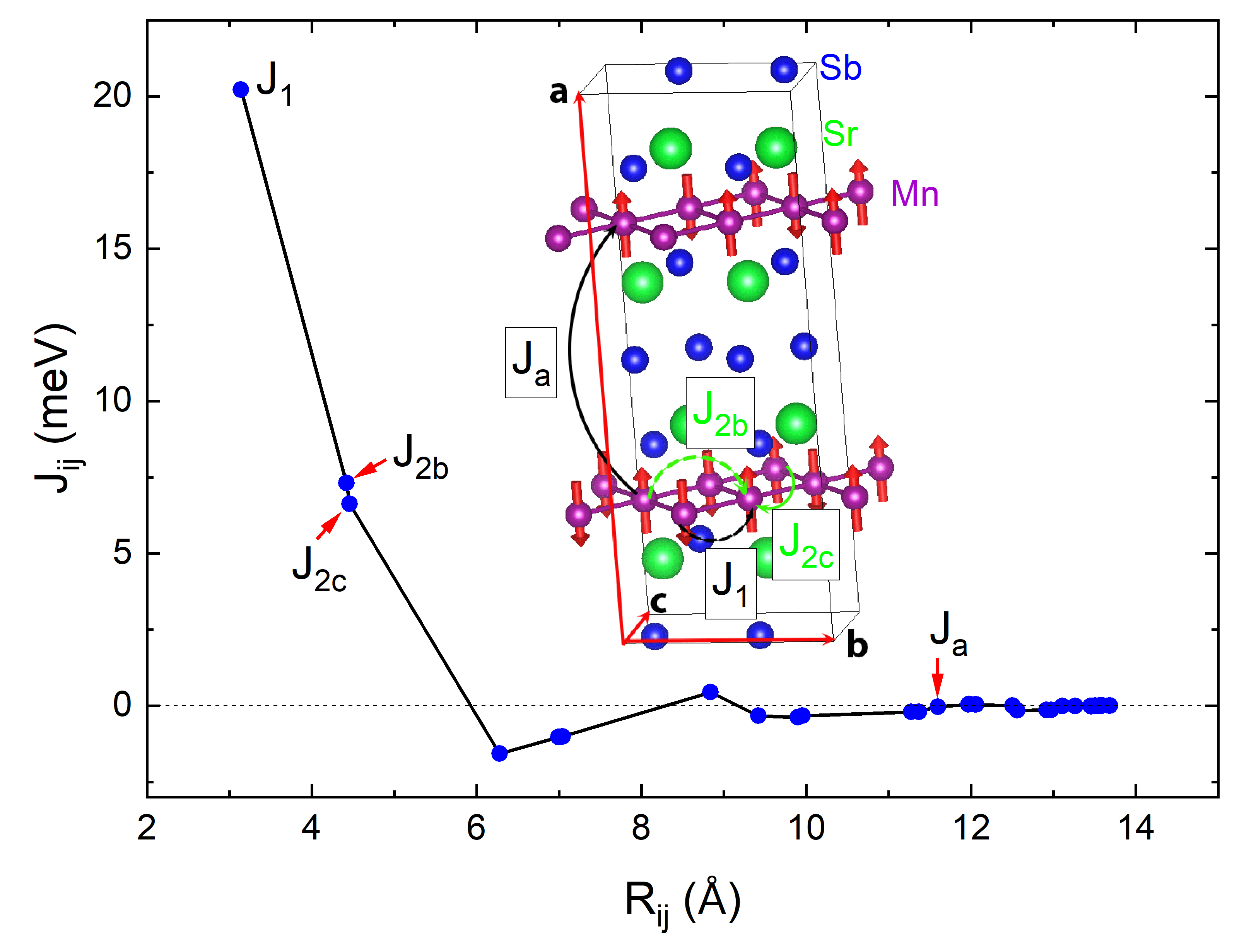}
\end{tabular}%
\caption{Real-space magnetic exchange parameters $J_{ij}$, defined as $J_{ij}^\text{N}$ in Eqs.~(\ref{eq:h-jn}) and (\ref{eq:jn-js}), in $\srmnsb$ as a function of neighbor distance $R_{ij}$.
  The spin configuration of the $C$-type magnetic ground state, NN exchange coupling $J_1$, NNN exchange couplings $J_{2b}$ and $J_{2c}$, and interlayer coupling $J_a$ are denoted in the inset.
}
\label{fig:j12cdist}
\end{figure}

\rFig{fig:j12cdist} presents the real-space magnetic exchange parameters $\jij$ as a function of Mn-Mn distance $R_{ij}$ obtained from the linear-response method.
It is dominated by NN exchange coupling $J_1$ and NNN exchange coupling $J_2$, showing a rapid decay as $R_{ij}$ increases and becoming negligible after $R_{ij}=\SI{8}{\AA}$.
The asymptotic behavior of fast decay suggests $\srmnsb$ a localized-moment system~\cite{ke2013prb}.
Both $J_1$ and $J_2$ exhibit positive values, indicating AFM interactions for both NN and NNN.
The amplitude of $J_2$ is about $1/3$ of $J_1$ and splits into two inequivalent couplings, $J_{2b}$ and $J_{2c}$, with $R_{ij}=$ \SI{4.42}{\AA} and \SI{4.46}{\AA}, respectively, due to the slight structural distortion.
Considering that the C-type configuration displays N\'{e}el-type in-plane AFM ordering, with AFM NN and FM NNN ordering within the $\bf bc$ basal plane, the sizable AFM $J_2$ suggests magnetic frustration.
%% The interlayer couplings with distances above $11\AA$ are much smaller than intralayer couplings. The calculated nearest interlayer coupling $J_a$ is ferromagnetic (FM), consistent with the C-type configuration. To incorporate other interlayer coupling contributions, we also extract the effective interlayer coupling using the energy mapping method, which also results in an FM interlayer coupling that agrees with experiments.

\begin{table}[htbp]
  \caption{Pairwise exchange parameters $\jij$ (\si{meV}) in $\srmnsb$ for the Heisenberg Hamiltonian $H=\sum_{i<j} J_{ij}\,\hat{\bf S}_i \cdot \hat{\bf S}_j$ as defined in \req{eq:h-jn}. 
  The  NN and NNN exchange parameters $J_1$, $J_{2b}$, and $J_{2c}$, and their corresponding distance $R_{ij}$ and Mn-$3d$ orbital resolutions are listed.
  As shown in ~\rfig{fig:crystalstructure}(a), we align the Mn $\bf bc$-basal plane in the $xy$ Cartesian plane for the convenience of orbital-contribution discussion.
  Calculated $SJ$ values are compared with corresponding INS values.
  }
\label{tbl:j12sym_orb}
\bgroup
\def\arraystretch{1.1}
\begin{tabular*}{\linewidth}{c@{\extracolsep{\fill}}cccccccccr}
\hline\hline  
$\jij$   & No. & $R_{ij}$  & $xy$ & $yz$ & $z^2$ &  $xz$ & $x^2-y^2$ & Total & $SJ$ & $SJ$(INS) \\ \hline
$J_1$    & 4   &  3.14  & 8.36 & 3.68 & 3.18 & 3.85 & 1.60 & 20.67 & 38.86 & 27.51(6)\\
$J_{2b}$ & 2   &  4.42  & 0.77 & 1.47 & 0.84 & 4.19 & 0.21 &  7.48 & 14.06 & 11.19(5)\\
$J_{2c}$ & 2   &  4.46  & 0.83 & 3.51 & 0.77 & 1.49 & 0.17 &  6.77 & 12.73 & 7.78(9)\\
\hline
\end{tabular*}
\egroup
\end{table}
\rtbl{tbl:j12sym_orb} summarizes the values of $J_1$, $J_{2b}$, and $J_{2c}$, along with their Mn-$3d$ orbital contributions, and corresponding values extracted from INS.
The dominant contribution to $J_1$ originates from the $d_{xy}$ orbital, which directly connects two NN Mn sites, as shown in \rfig{fig:crystalstructure}(c).
Other $d$ orbitals make smaller AFM contributions to $J_1$.
Additionally, the $d_{yz}$ and $d_{xz}$ orbitals exhibit slightly different contributions, reflecting the small distortion in the Mn basal plane. 

$J_{2b}$ and $J_{2c}$ are oriented along the $\mathbf{b}$ ($\hat{x}$) and $\mathbf{c}$ ($\hat{y}$) directions, respectively.
Consequently, their primary contributions arise from $d_{xz}$ and $d_{yz}$ orbitals, respectively.
An intriguing observation is that the $d_{x^2-y^2}$ orbital, which aligns directly with the coupling directions of $J_{2b}$ and $J_{2c}$, contributes the least.
This suggests that the direct exchange is relatively small due to the greater distance associated with $J_{2}$ and stands in stark contrast to the large contribution of $d_{xy}$ to $J_1$.
Taken together, it becomes evident that the indirect superexchange through the Sb layers adjacent to the Mn layer plays a significant role in determining the magnitude of $J_2$.
Conversely, $J_{2b}$ exhibits a slightly larger value than $J_{2c}$, possibly attributable to the former's shorter bond length compared to the latter, resulting from the subtle structural distortion.

The interlayer exchange couplings are notably weaker owing to the substantial distance ($R_{ij}>\SI{11}{\AA}$) between Mn layers along the $\mathbf{a}$ ($\hat{z}$) direction.
The calculated nearest interlayer coupling $J_a$ is FM, consistent with the experimental C-type spin ordering, and has a value of \SI{-0.029}{meV}, which is comparable with INS measurements as we discuss later.

As shown in \rtbl{tbl:j12sym_orb}, in comparison to INS measurements, calculations somewhat overestimate the $SJ$ values.
Nonetheless, it is essential to highlight that the calculated trend in $SJ$ values aligns with the experimental trend.
This underscores the consistency between the theoretical predictions and experimental observations.
The exchange interactions are also calculated by mapping multiple magnetic configurations to a minimalistic $J_1$-$J_2$-$J_a$ Heisenberg model (See details in Appendix \ref{Jij_fittingE}).
We found $J_1^\text{eff}=\SI{22.47}{\meV}$, $J_2^\text{eff}=\SI{5.25}{\meV}$, and $J_a^\text{eff}=\SI{-0.61}{\meV}$.
Both $J_1^\text{eff}$ and $J_2^\text{eff}$ show good agreement with INS results and our linear response calculations.
The values of $J_1^\text{eff}$ and $J_a^\text{eff}$ are consistent with those reported in a prior DFT study~\cite{zhang2019prb}, whereas $J_a^\text{eff}$ is approximately one order of magnitude larger than both the INS results and our linear response calculations.
Overall, as we demonstrate later, the values of $J_1$, $J_2$, and $J_2/J_1$ calculated using the linear response theory exhibit better agreement with experimental data when compared to the mapping method.

\begin{figure}[ht]
\centering
\begin{tabular}{c}
\includegraphics[width=.8\linewidth,clip]{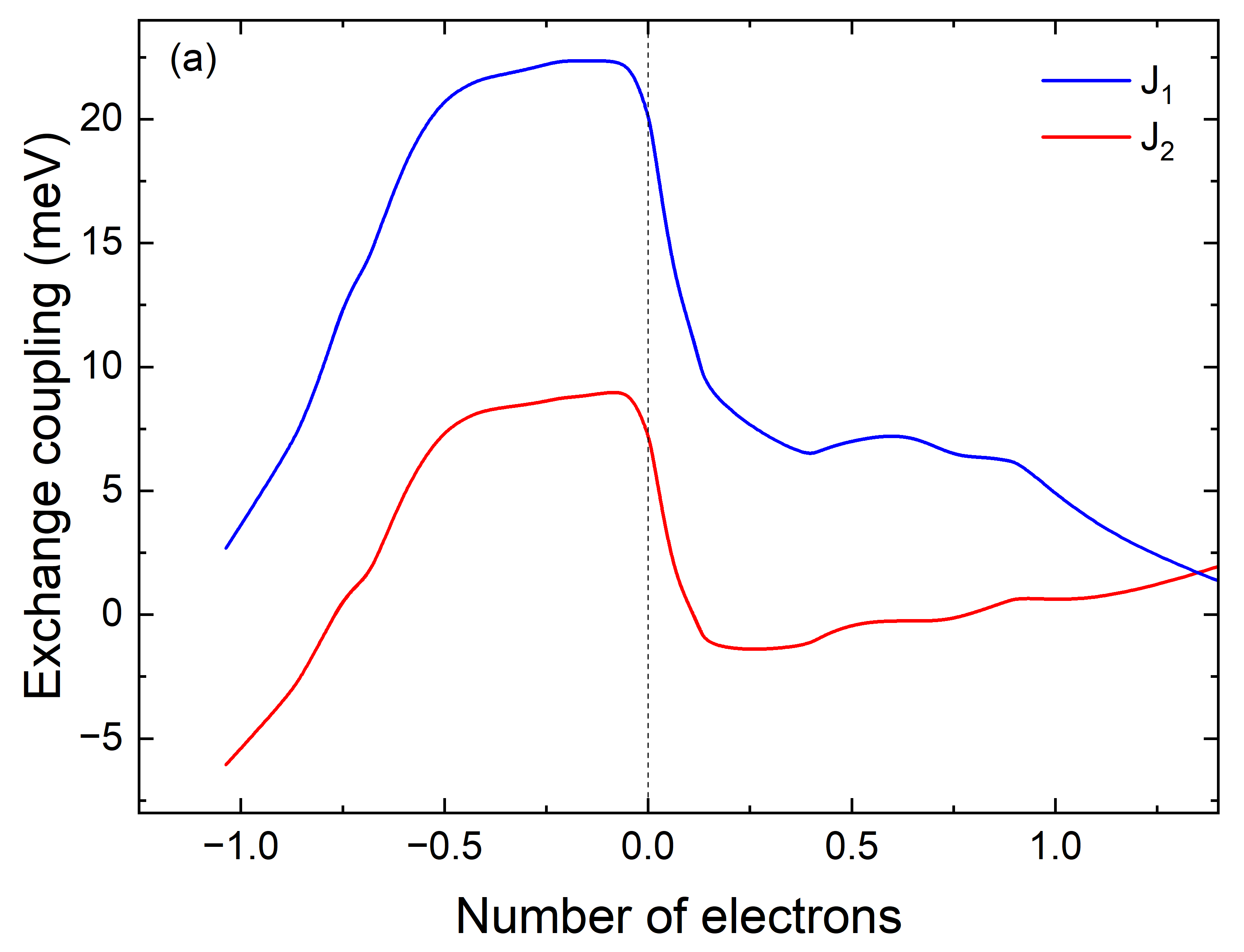}   \\
\includegraphics[width=.8\linewidth,clip]{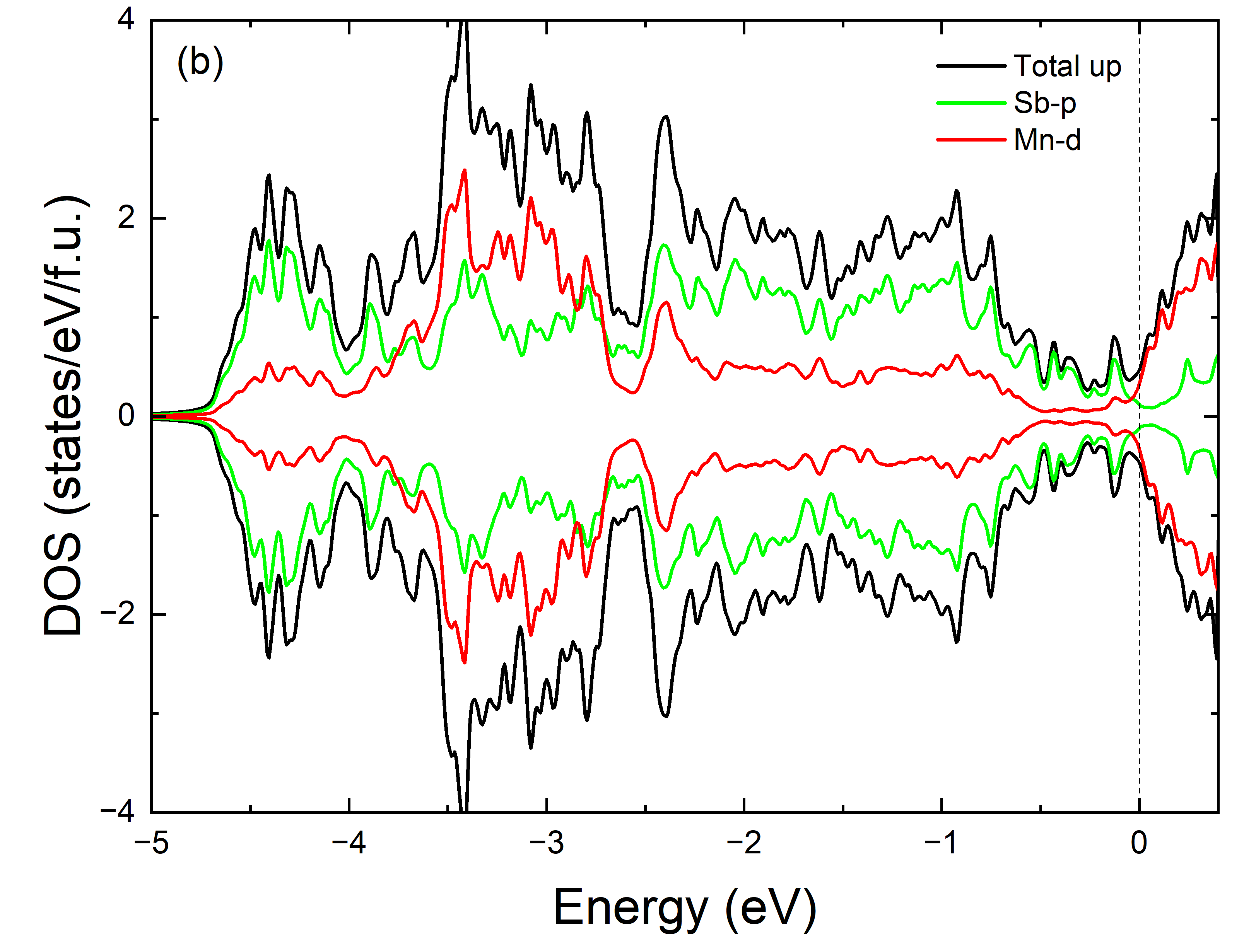} \\  
\includegraphics[width=.8\linewidth,clip]{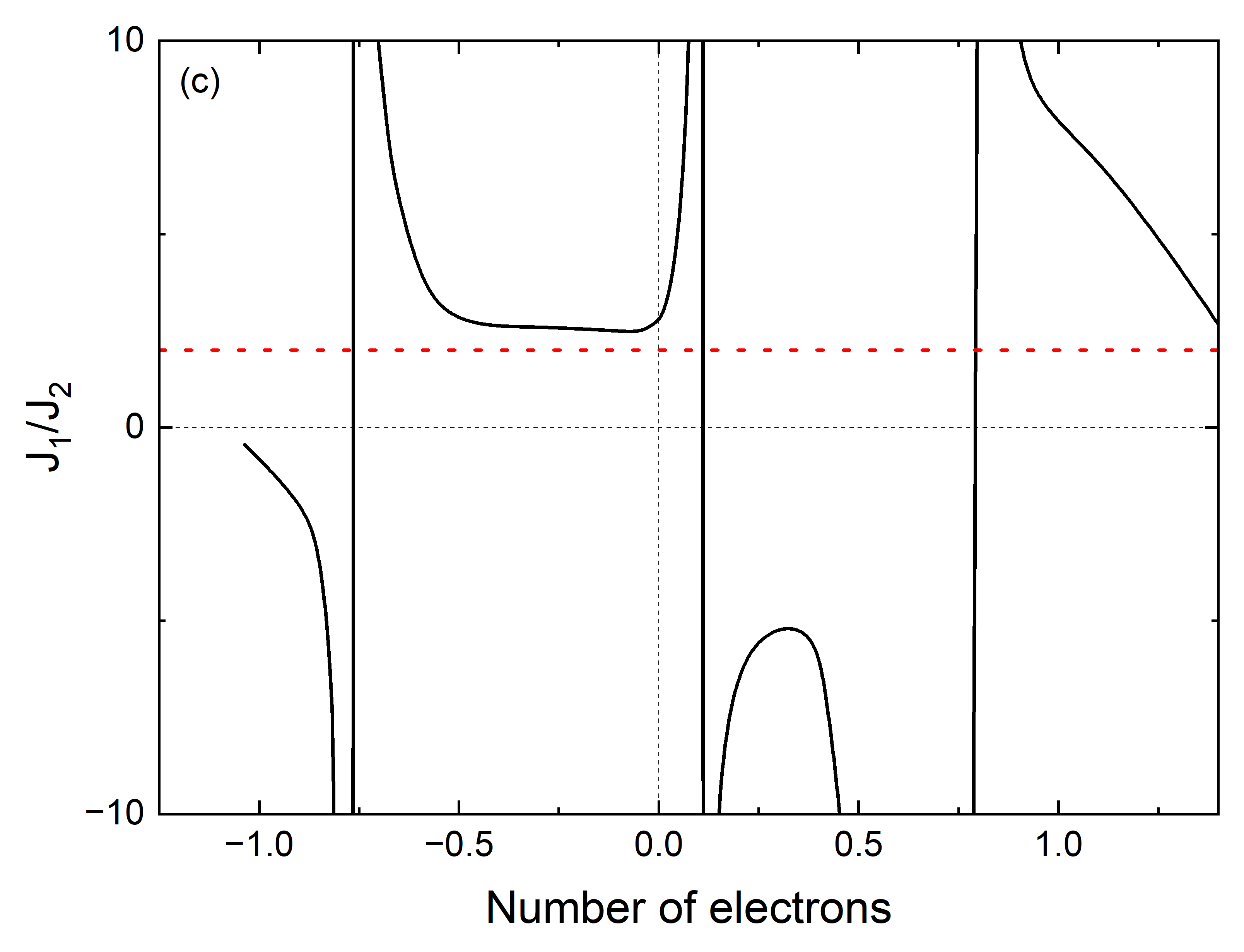}
\end{tabular}%
\caption{Bandfilling dependence of exchange parameters $J_1$ and $J_2$, and density of states (DOS) in $\srmnsb$ calculated in TB.
The Fermi level is shifted to zero.
(a) $J_1$ and $J_2$ as functions of bandfilling (Number of doping electrons per f.u.).
(b) The total DOS, Sb-$p$, and Mn-$d$ partial DOS (Number of states per f.u. per eV).
(c) The ratio of $J_1/J_2$ as a function of doping, where the horizontal red dashed line at $J_1/J_2=2$ corresponds to maximum frustration.
}
\label{fig:j12cw}
\end{figure}
To explore the effects of electron or hole doping on exchange couplings and magnetic ordering, we calculate $J_{ij}$ as functions of band filling within a rigid-band approximation.
\rFig{fig:j12cw}(a) illustrates the dependence of the exchange parameters $J_1$ and $J_2$ on the number of doping electrons. (The corresponding orbital resolution can be found in the appendix.)
With hole doping, both $J_1$ and $J_2$ increase slightly before decreasing.
The energy range spanning from \SI{-0.5}{\eV} to $\ef$ is primarily governed by Sb-$5p$ states in the density of states (DOS), with Mn-$3d$ states contributing minimally, as depicted in \rFig{fig:j12cw}(b).
Consequently, the values of $J_1$ and $J_2$ exhibit only marginal alterations in the range of 0.1--0.45 hole/f.u.
doping.
The ratio of $J_1$ and $J_2$ is shown in \rfig{fig:j12cw}(c), where the red dashed lines indicate the points where the square lattice reaches maximum frustration, residing on the boundary between N\'{e}el- and stripe-type AFM configurations within the basal plane.
The $J_1/J_2$ ratio does not change significantly with up to 0.6 hole/f.u doping.
Furthermore, the resulting $J_0^e$ increases slightly with weak hole doping, indicating a small increase in $T_\text{N}$.
These computational results align well with a previous experimental study on hole doping~\cite{Liu2019prbhd}, which found that the magnetic structure remains unchanged, maintaining the same in-plane N\'{e}el-type magnetic ordering with pronounced frustration, albeit with a slight increase in $T_\text{N}$.

Electron doping, on the other hand, has a much stronger effect on $J_1$ and $J_2$; doping within 0.1 electron/f.u.
rapidly reduces their AFM coupling strength, with $J_2$ even transitioning to a weakly FM interaction.
The resulting increase in the $J_1/J_2$ ratio indicates weaker spin frustration.

\begin{figure}[ht]
	\centering
	\includegraphics[width=.80\linewidth,clip]{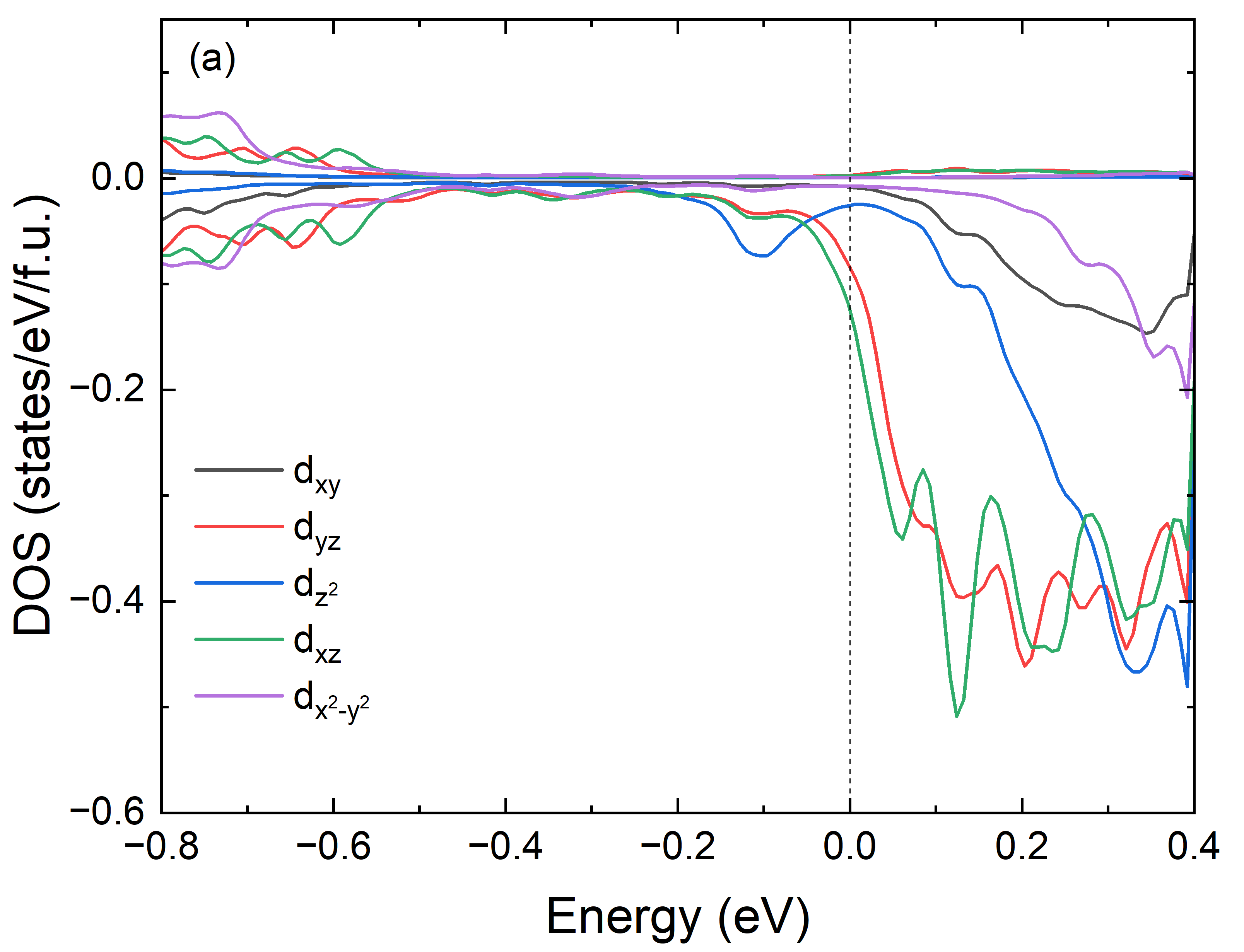}
	\includegraphics[width=.80\linewidth,clip]{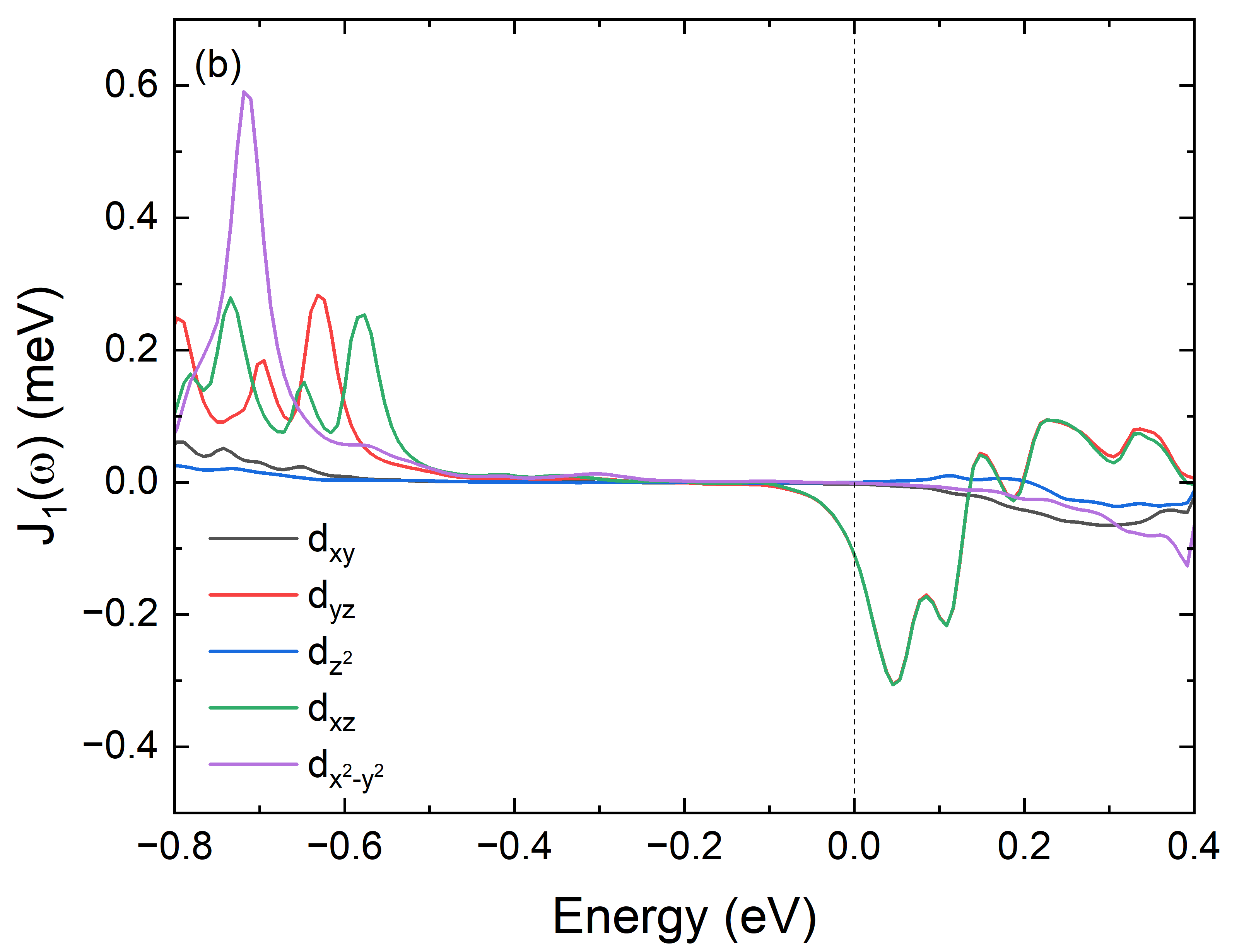}
	\includegraphics[width=.80\linewidth,clip]{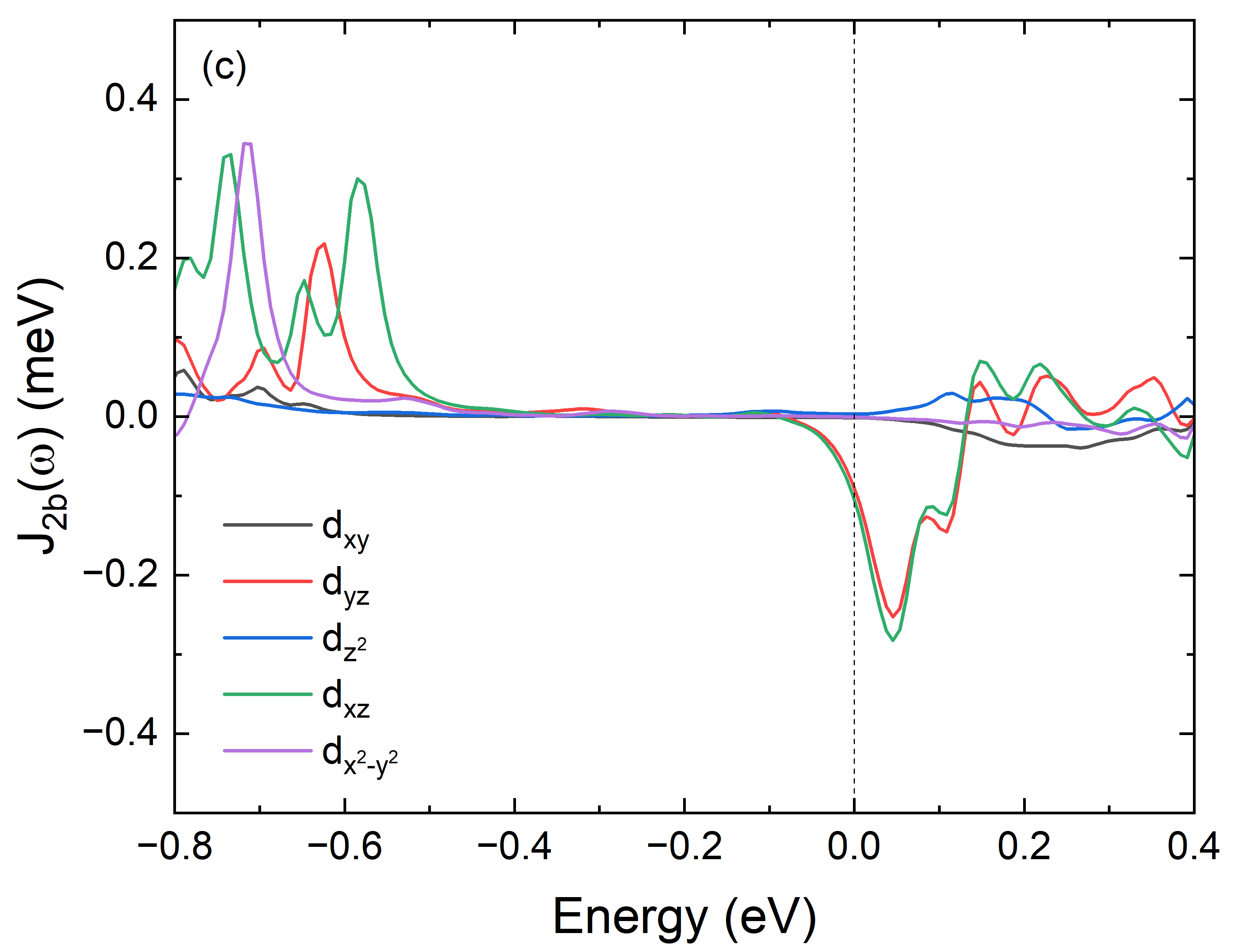}
	\caption{(a) The partial DOS projected onto $3d$ orbitals of a single Mn site and orbital- and bandfilling-resolved exchange couplings: (a) $J_1$ and (b) $J_{2b}$. The bandfilling-resolved exchange couplings, such as $J_1(\omega)$, are defined as $J_1 = \int_{-\infty}^{\ef} \ud\omega J_1(\omega)$. $J_{2c}(\omega)$ behaves similarly to $J_{2b}(\omega)$ and is therefore not shown.
        }
	\label{fig:DOS_Mn_J1w_J2w}
\end{figure}

To understand the simultaneous decrease of $J_1$ and $J_{2b}$, we further resolve these parameters into orbitals and bandfilling contributions.
\rFig{fig:DOS_Mn_J1w_J2w} shows the resulting resolved $J_1(\omega)$ and $J_{2b}(\omega)$ along with the partial DOS projected on Mn-$3d$ orbitals.
Here, the bandfilling-resolved exchange couplings, $J_{ij}(\omega)$, are defined as $J_{ij} = \int_{-\infty}^{\ef} \ud\omega J_{ij}(\omega)$, analogous to the density of states $D(E)$ and the total number of electrons $N=\int_{-\infty}^{\ef} \ud\omega D(\omega)$.

As depicted in \rfig{fig:DOS_Mn_J1w_J2w}(a), the energy window of \SIrange{-0.4}{0.4}{\eV} exhibits negligible DOS in the majority-spin channel, with $\ef$ positioned at the shoulder of Mn-$d_{xz|yz}$ states in the minority spin channel.
As shown in \rfig{fig:DOS_Mn_J1w_J2w}(b) and (c), in the energy range of \SIrange{-0.2}{0.2}{\eV} near $\ef$, all orbital-resolved $J_1(\omega)$ and $J_{2b}(\omega)$ values are mostly negative, indicating contributions to FM interactions.
Conversely, for occupied states ranging from \SIrange{0.6}{0.8}{\eV} below $\ef$, we have $J_1(\omega)>0$ and $J_{2b}(\omega)>0$, suggesting that the occupied states in this range contribute to AFM interactions for both $J_1$ and $J_{2b}$.

Therefore, in the rigid-band model, a slight hole doping (shifting $\ef$ down to approximately \SI{0.4}{\eV} below $\ef$) would decrease the FM contribution, making $J_1$ and $J_{2}$ more AFM-like.
In contrast, both $J_1$ and $J_{2b}$ become less AFM (or more FM) with further hole doping, which removes the AFM contribution from states in the energy window of \SIrange{0.5}{0.8}{\eV} below $\ef$, or with electron doping that adds FM contribution from states in the energy window of \SIrange{0}{0.2}{\eV} above $\ef$.
Notably, the bandfilling dependence of both $J_1$ and $J_{2b}$ primarily correlates with the filling of $d_{xz}$ and $d_{yz}$ states near $\ef$.
Moreover, unsurprisingly, despite a substantial increase in the partial DOS above $\ef$ for the out-of-plane Mn-$d_{z^2}$ orbitals, its contribution to in-plane couplings is minimal.

\subsection{Spin wave spectra in $J_1$-$J_{2b}$-$J_{2c}$-$J_a$ model}

Starting from the C-type ground state, we can derive the magnon dispersions of the $J_1$-$J_{2b}$-$J_{2c}$-$J_a$ model for $\srmnsb$ using the LSWT.
The energies of the two magnon bands at $\mathbf{k}= H\mathbf{a}^*+ K\mathbf{b}^*+ L\mathbf{c}^*$ can be written as:
\begin{eqnarray}
  \omega(\bfk) &=& M_0 \sqrt{(\a_k-d_0\pm |\zeta_k|)^2 -|\b_k|^2}\,,
\label{sweig5}
\end{eqnarray}
where $M_0$ is the amplitude of the on-site Mn moment calculated in DFT, and
\begin{eqnarray}
\a_k &=& J_{2b}\cos(2\pi K) + J_{2c}\cos(2\pi L) \,, \\ \nonumber
\b_k &=& \frac12 J_{1}\left[1+ e^{i2\pi( K + L)} + e^{i2\pi K}  + e^{i2\pi L} \right] \,, \\  \nonumber
\zeta_k &=& \frac12 J_a (1+e^{i2\pi H}) \,, \\ \nonumber
d_0 &=& J_{2b} + J_{2c} + J_{a}-2J_1 \,.
\end{eqnarray}

\begin{figure}[ht]
\centering
\begin{tabular}{c}
\includegraphics[width=0.99\linewidth,clip]{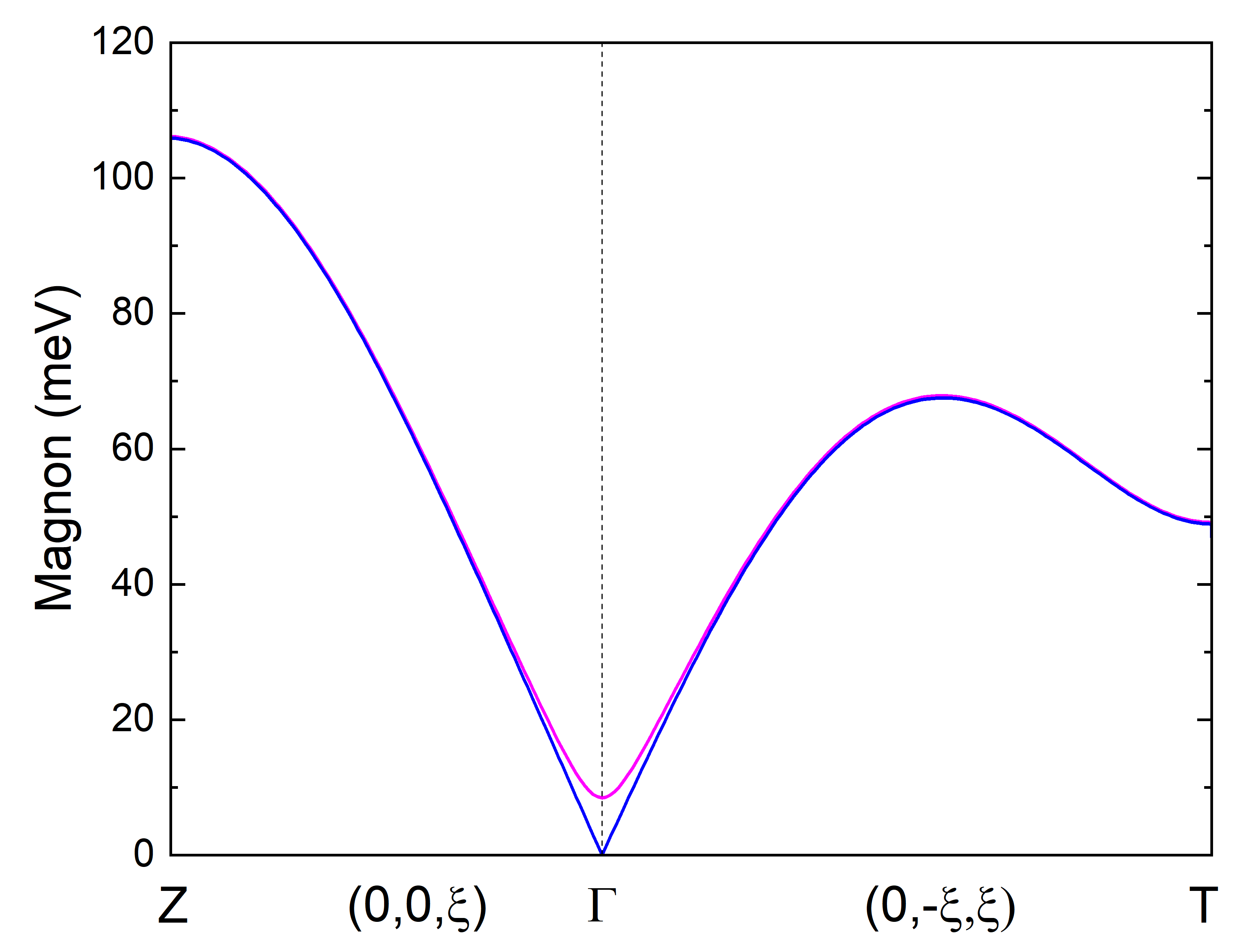} 
\end{tabular}%
\caption{ Magnon band structure in $\srmnsb$ calculated using the $J_1$-$J_{2b}$-$J_{2c}$-$J_a$ model with parameters obtained from DFT.
Magnetocrystalline anisotropy is not included.
High-symmetry $k$ points Y and T are illustrated in \rfig{fig:crystalstructure}(d).}
\label{fig:spinwave}
\end{figure}
\rFig{fig:spinwave} shows the magnon band structure along the Z--$\Gamma$--T path calculated using \req{sweig5} with intralayer coupling parameters listed in \rtbl{tbl:j12sym_orb} and interlayer coupling $J_a$.
It is worth noting that $\Gamma$--Z [$(0,0,\xi)$] corresponds to the $J_{2c}$ direction, while $\Gamma$--T [$(0,-\xi,\xi)$] is slightly deviated from the $J_1$ direction.
The inclusion of interlayer coupling $J_a$ splits the magnon bands at $\Gamma$, as the acoustic and optical modes corresponding to the in- and out-of-phase spin precession with respect to the two Mn layers in the unit cell become energetically distinguishable.
Although the calculation slightly overestimates the magnon bandwidth, the dispersions are in overall good agreement with INS measurements described below.

%% A detailed comparison of the values of $J_1$ obtained using both symmetrized and unsymmetrized Hamiltonians, as well as spin wave spectra calculated along high-symmetry directions, can be found in Section I of the supplemental material~\cite{suppl_srmnsb2}.

\subsection{INS measurements and simulations}

\begin{figure}[ht]
	\centering
	\includegraphics[width=1.0\linewidth,clip]{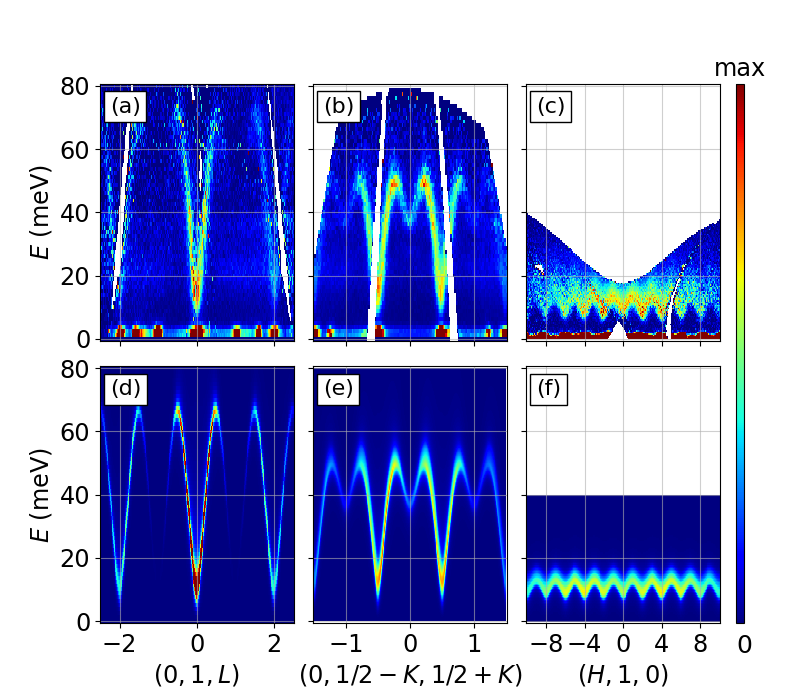}
	\caption{INS data (a-c) and simulations using experimental parameters (d-f) of the SW spectra along three high-symmetry directions $[0,1,L]$ ($\Gamma$--Z), $[0,1/2-K,1/2+K]$, and $[H,1,0]$ ($\Gamma$--X).
	}
	\label{fig:INS}
\end{figure}

\begin{figure}[ht]
	\centering
	\includegraphics[width=1.0\linewidth,clip]{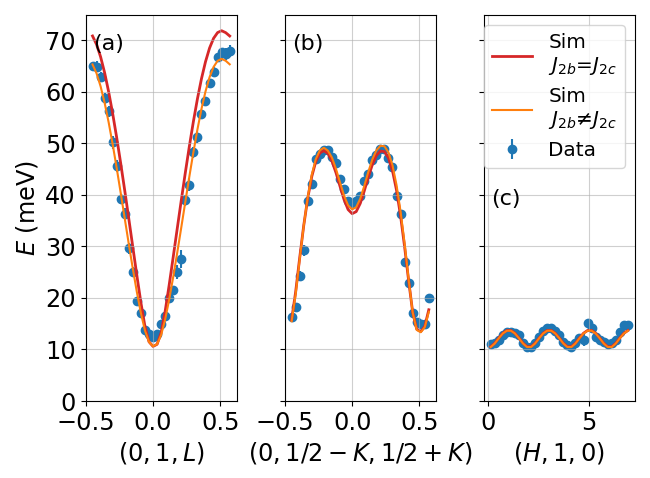}
	\caption{Magnon dispersion along three high-symmetry directions determined from fits to the INS data.  
	Lines show fits to the dispersion using the two different model Hamiltonians discussed in the text.
	Model 1 with $J_{2b}=J_{2c}$ is shown as red solid line while model 2 with $J_{2b} \neq J_{2c}$ is shown as orange solid line.
	}
	\label{fig:INS_disp}
\end{figure}

\rFig{fig:INS} shows the INS data (a--c) and simulations (d--f) of the SW spectra along three high-symmetry directions $(0,1,L)$ ( $\Gamma$--Z ), $(0,1/2-K,1/2+K)$ ( $\Gamma$--T ), and $(H,1,0)$ ( $\Gamma$--X). 
The simulations are done through experimentally determined parameters as presented in \rtbl{tbl:j12sym_orb} (or Model 2 in \rtbl{tbl:INS}).
Constant energy cuts of the data were taken and peaks in the cuts were fits to Gaussian functions in order to determine the peak center for a given $E$.
The extracted peak centers of the energy line cuts are presented in Fig.~\ref{fig:INS_disp}. 
The data treatment and modelling and the cuts and fits are given in Section II of the supplemental material~\cite{suppl_srmnsb2}.

\begin{table}[bhtp]
	\caption{Parameters and single-ion anisotropy ($SD$) and magnetic interactions (\si{meV}) from two models.
	}
	\label{tbl:INS}%
	\begin{tabular*}{\linewidth}{l @{\extracolsep{\fill}} rrrrrrr}
	\hline\hline
		 Model & $SD$  & $SJ_1$ & $SJ_{2b}$  & $SJ_{2c}$ & $SJ_{a}$ & r$\chi^2$ \\ 
		 \hline
		1 & -0.26(6)  & 27.51(6) &   9.36(2)   &  9.36(2) &  -0.092(6) & 15.04 \\  
		2 & -0.2559(9)  & 27.96(1) &   11.19(5)   &  7.78(9) &  -0.091(3) & 4.79 \\
		\\[-1.1em]
		\hline
	\end{tabular*}
\end{table}

To understand the INS data, we calculate the SW spectra using LSWT with two models: one with (Model 1) and the other without (Model 2) enforcing $J_{2b}=J_{2c}$.
The fitted single-ion anisotropy $SD$, in-plane magnetic interactions $J_1$, $J_{2b}$, $J_{2c}$, and inter-layer interaction $J_{a}$ are summarized in \rtbl{tbl:INS}.

In Model 2, the difference between $J_{2b}$ and $J_{2c}$ induces asymmetry between the magnon bands along the path $(0,1/2-K,1/2+K)$, as shown in \rfig{fig:INS_disp}(b), resulting in higher energy for positive $K$ values than negative $K$ values.
Additionally, the observed structural twins introduce a mixing of $J_{2b}$ and $J_{2c}$.
Due to the small difference between $J_{2b}$ and $J_{2c}$ and the relatively coarse energy resolution used for our measurements, the splitting is not observed in the INS data.
Nonetheless, Model 2 provides a slightly better fit to the dispersion with a smaller reduced $\chi^2$ (r$\chi^2$) compared to Model 1, as shown in \rfig{fig:INS_disp} and \rtbl{tbl:INS}.

Overall, when compared to INS measurements, the SW dispersion derived from first-principles calculations, although somewhat overestimating the bandwidth, agrees reasonably well with INS measurements.
Correspondingly, the intralayer interactions ($J_1$, $J_{2b}$, and $J_{2c}$) as well as the interlayer interaction ($J_a$) are comparable to values extracted from INS.
The concordance between the theoretical predictions and the experimentally-derived $SJ$-based calculations affirms the accuracy of our first-principles methodology in effectively predicting magnetic interactions within the materials.

The overestimation of $SJ$ values in our calculations could potentially be addressed.
For instance, the MLWFs we used here typically have tails extending into other sites, which may blur the definition of calculated exchange couplings between two sites.
It would be worthwhile to investigate whether a representation of the on-site moment in a basis that better localizes the magnetic moment within the atomic sphere could improve the description.
Another consideration is the inclusion of electron-correlation effects beyond DFT.
Typically, one may consider additional electron repulsion for the localized Mn-$3d$ orbitals beyond DFT using methods such as DFT+$U$.
This approach often promotes electron localization, thereby increasing the magnetic moment while decreasing the exchange coupling.
Our preliminary results from applying an additional Hubbard-like $U$ potential in DFT to Mn-$3d$ orbitals show enhanced electron localization and reduced intersite exchange couplings, leading to a better agreement with experiments in terms of overall magnon bandwidth.
However, ideally, the choice of the $U$ values should be justified, for example, by comparing calculated and experimentally measured band structures.
$GW$-based methods can be even more useful, as they are not only parameter-free but also contain nonlocal off-site exchange-correlations that do not exist in DFT+$U$ but can be crucial for properly describing the superexchanges.
A systematic study of electron correlation effects is beyond the scope of the present work.

\section{Conclusions}
This study on $\srmnsb$ presents a comprehensive understanding of its electronic and magnetic properties, achieved through a synergy of theoretical methodologies and experimental data.
The exploration of exchange coupling parameters in $\srmnsb$ is carried out using linear response theory within a realistic TB model.
Our calculated intralayer interactions ($J_1$, $J_{2b}$, and $J_{2c}$) and interlayer interaction ($J_a$) are in good agreement with INS results.
Moreover, the study reveals a sizable AFM NNN exchange coupling $J_2$, introducing significant spin frustration within the basal plane.
Orbital-resolved contributions to exchange couplings help reveal the microscopic origin of the exchange interactions.
Notably, we find that the NN exchange coupling $J_1$ in $\srmnsb$ predominantly arises from the contributions of Mn $d_{xy}$ orbitals, aligning with the nearest Mn-Mn bond, while $J_{2b}$ and $J_{2c}$ receive primary contributions from $d_{xz}$ and $d_{yz}$ orbitals, respectively.
This reflects the dominance of direct-exchange and superexchange nature for $J_1$ and $J_2$, respectively.
The band-filling dependence of exchange coupling based on the rigid-band model reveals that electron doping is expected to weaken both $J_1$ and $J_2$ while relieving spin frustration through increasing the $J_1/J_2$ ratio.
Moreover, the magnetic structure of $\srmnsb$ is anticipated to remain unaltered under carrier doping, which is in agreement with previous hole-doping investigations~\cite{Liu2019prbhd}.
Introducing additional electron correlation within the Mn-$3d$ orbitals can promote electron localization and reduce the magnetic coupling, further improving the agreement with experiments.
In summary, this investigation not only advances our comprehension of $\srmnsb$ but also underscores the efficacy of combining \emph{ab initio} and experimental methods in elucidating intricate materials and their magnetic interactions.

\section{Acknowledgments}
This work is supported by the U.S.~Department of Energy (DOE), Office of Basic Energy Sciences, Division of Materials Sciences and Engineering.
WT acknowledges support from the DOE Science Undergraduate Laboratory Internships (SULI) program.
Ames Laboratory is operated for the U.S.
Department of Energy by Iowa State University under Contract No.~DE-AC02-07CH11358.
This research used resources at the Spallation Neutron Source, a DOE Office of Science User Facility operated by the Oak Ridge National Laboratory.

\appendix

\section{Hamiltonian Symmetrization}\label{symham}
Wannier functions (WFs) are localized atomic-like orbitals defined at each atom site.
For a set of isolated bands that are separated from all other lower and higher bands throughout the BZ by band gaps, its electronic states can be well described by a set of WFs $|w_{\mathbf{R}}^{\mu}\rangle$, where $\mathbf{R}$ is the location of the unit cell considered, and $\mu$ is the index of WFs in the cell.
For a multiple-site system, $\mu$ is a combination of quantum numbers $n_i$, $l_i$, $m_i$, $\sigma_i$, and site position vectors $\boldsymbol\tau_i$.
For systems with only one site in the primitive cell, we have $\boldsymbol\tau_i=(0,0,0)$.

When a WF is rotated by one point symmetry operation $\hat{\mathcal{R}}$ of the system, it can only transform into WFs belong to equivalent sites with the same quantum numbers $n_i$ and $l_i$, that is, the symmetry can only mix states with different magnetic quantum numbers and spins
\begin{eqnarray}
  \hat{\mathcal{R}} |w_{\mathbf{R}+\boldsymbol\tau_i}^{ \b_i m_i \sigma_i}\rangle = \sum_{m'_i \sigma'_i} \mathcal{R}^{l_i}_{m'_i\sigma_i'm_i\sigma_i} |w_{\mathbf{R}'+\boldsymbol\tau'_i}^{ \b_i m'_i \sigma'_i}\rangle \,,
\end{eqnarray}
where $\hat{\mathcal{R}}(\mathbf{R}+\boldsymbol{\tau}_i)=\mathbf{R}'+\boldsymbol{\tau}'_i$ and $\b_i$ the combination of $n_i$ and $l_i$. 
According to the orthogonalization of WFs, the transformation matrix $\mathcal{R}_{\mathbf{R}'\mathbf{R}}$ with elements given by
\begin{align}
  \langle w_{\mathbf{R}'+\boldsymbol\tau'_i}^{ \b_i m'_i \sigma'_i} |\hat{\mathcal{R}} |w_{\mathbf{R}+\boldsymbol\tau_i}^{ \b_i m_i \sigma_i}\rangle=\mathcal{R}^{l_i}_{m'_i\sigma'_i,m_i\sigma_i} \d_{\mathbf{R}'+\boldsymbol{\tau}'_i,\hat{\mathcal{R}}(\mathbf{R}+\boldsymbol{\tau}_i)} \delta_{\b_i,\b'_i}\,,
\end{align}
It is obvious that the transformation matrix $\mathcal{R}$ should be $l$ and $n$ diagonal. 

The transformation matrix $\mathcal{R}$ can be obtained by the behavior of \textsc{Wannier90}'s basis orbitals under symmetry operations.
\textsc{Wannier90} uses real spherical harmonics $\mathcal{Y}_{l,m}$ which is related to complex spherical harmonics $Y_{l}^{m}$ by
\begin{eqnarray}\label{ctor}
  \mathcal{Y}_{l,m}&=& \begin{cases}
    \sqrt{2}\,(-1)^m\Re\{Y_{l}^{|m|}(\theta,\phi)\} & m>0 \\
    \sqrt{2}\,(-1)^m\Im\{Y_{l}^{-|m|}(\theta,\phi)\} & m<0 \\
    \Re\{Y_{lm}(\theta,\phi)\} & m=0
  \end{cases} \,, \nn
  &=&\begin{cases}\frac{1}{\sqrt{2}}\left[  Y_{l}^{-|m|}+ (-1)^m Y_{l}^{|m|}\right]  & m>0 \\
    \frac{i}{\sqrt{2}}\left[ Y_{l}^{-|m|}- (-1)^m Y_{l}^{|m|}\right]  & m<0 \\
    Y_{l}^0 & m=0 \end{cases}\,.
\end{eqnarray}
This equation defines the unitary transformation matrix $\bold{U}_{\mathbb{C}\rightarrow \mathbb{R}}$ between $\mathcal{Y}_{l,m}$ and $Y_{l}^{m}$.
The angular momentum operator $\hat{L}$ in the basis of real spherical harmonics can then be written as
\begin{eqnarray}
  \hat{L}^{\mathbb{R}}=\bold{U}_{\mathbb{C}\rightarrow \mathbb{R}} \hat{L}^{\mathbb{C}} \bold{U}_{\mathbb{C}\rightarrow \mathbb{R}}^{\dagger}=\bold{U}_{\mathbb{C}\rightarrow \mathbb{R}} \hat{L}^{\mathbb{C}} \bold{U}_{\mathbb{R}\rightarrow \mathbb{C}} \,.
\end{eqnarray}
In the presentation of complex spherical harmonics $Y_{\ell }^{m}$, the non-vanished matrix elements of $\hat{L}^{\mathbb{C}}$ are
\begin{eqnarray}
  \langle l,m   | \hat{L}_z |l,m \rangle &=& \hbar m \nonumber, \\ 
  \langle l,m-1 | \hat{L}_- |l,m \rangle &=& \hbar \sqrt{l(l+1)-m(m-1)},  \\
  \langle l,m+1 | \hat{L}_+ |l,m \rangle &=& \hbar \sqrt{l(l+1)-m(m+1)}.  \nonumber
\end{eqnarray} 
The rotation on orbital function is defined by
\begin{eqnarray}
  \hat{\mathcal{R}}^{\mathbb{C}}=e^{-i\hat{\mathbf{n}}\cdot \hat{L}^{\mathbb{C}}\phi} \,,
\end{eqnarray} 
where $\hat{\mathbf{n}}$ is the axis of the rotation symmetry and $\phi$ the rotation angle. Then the transformation matrix $\mathcal{R}$ for the orbital part can be obtained by
\begin{eqnarray}
  \hat{\mathcal{R}}^{\mathbb{R}}=\bold{U}_{\mathbb{C}\rightarrow \mathbb{R}} \hat{\mathcal{R}}^{\mathbb{C}} \bold{U}_{\mathbb{R}\rightarrow \mathbb{C}}=\bold{U}_{\mathbb{C}\rightarrow \mathbb{R}} e^{-i\hat{\mathbf{n}}\cdot \hat{L}^{\mathbb{C}}\phi} \bold{U}_{\mathbb{R}\rightarrow \mathbb{C}}\,.
\end{eqnarray}
The rotation on spinor can be written as
\begin{eqnarray}
  \hat{\mathcal{R}}^{\hat{\sigma}}=e^{-i\hat{\mathbf{n}}\cdot \hat{S}\phi} \,.
\end{eqnarray}
The spin angular momentum operator $\hat{S}$ is defined as
\begin{eqnarray}
  \hat{S} = \dfrac{\hbar}{2} (\sigma_x, \sigma_y, \sigma_z) \,,
\end{eqnarray}
where $\sigma_x$, $\sigma_y$ and $\sigma_z$ are Pauli matrices given by
\begin{eqnarray}
  \sigma_x = \begin{pmatrix}
    0 & 1 \\
    1 & 0
  \end{pmatrix} \,, \quad
  \sigma_y = \begin{pmatrix}
    0 & -i \\
    i & 0
  \end{pmatrix} \,, \quad
  \sigma_z = \begin{pmatrix}
    1 & 0 \\
    0 & -1
  \end{pmatrix} \,.
\end{eqnarray}
Thus, in the representation of real spherical harmonics and spinor, the rotation is given by $\hat{\mathcal{R}} = \hat{\mathcal{R}}^{\mathbb{R}} \hat{\mathcal{R}}^{\hat{\sigma}} $.

The Hamiltonian $\hat{H}$ of a system is invariant under its all symmetry operations
\begin{eqnarray}
  \hat{H}=\hat{\mathcal{R}} \hat{H} \hat{\mathcal{R}}^{\dagger}\,.
\end{eqnarray}	
The elements of real-space Hamiltonian $H^{\mathbf{R}}$ are defined by $H^{\mathbf{R}}_{ij}=\langle w_{\boldsymbol\tau_i}^{ \b_i m_i \sigma_i} |\hat{H} |w_{\mathbf{R}+\boldsymbol\tau_j}^{\b_j m_j \sigma_j}\rangle$ that under symmetry operations transforms as
\begin{align}
  H^{\mathbf{R}}_{ij}&=\langle w_{\boldsymbol\tau_i}^{ \b_i m_i \sigma_i} |\hat{H} |w_{\mathbf{R}+\boldsymbol\tau_j}^{ \b_j m_j \sigma_j}\rangle \nn
  &= \langle w_{\boldsymbol\tau_i }^{\b_i m_i \sigma_i} |\hat{\mathcal{R}} \hat{H} \hat{\mathcal{R}}^{\dagger} |w_{\mathbf{R}+\boldsymbol\tau_j}^{ \b_j m_j \sigma_j}\rangle\nn
  &=\sum_{\mu_i' \mu_j',\mathbf{R}' \mathbf{R}^{\prime\prime}} \langle w_{\boldsymbol\tau_i}^{ \b_i m_i \sigma_i} |\hat{\mathcal{R}} | w_{\mathbf{R}'+\boldsymbol\tau'_i}^{ \b'_i m'_i \sigma'_i} \rangle \langle w_{\mathbf{R}'+\boldsymbol\tau'_i}^{ \b'_i m'_i \sigma'_i}  |\hat{H}| w_{\mathbf{R}^{\prime\prime}+\boldsymbol\tau'_j}^{ \b'_j m'_j \sigma'_j} \rangle \nn
  &\times \langle w_{\mathbf{R}^{\prime\prime}+\boldsymbol\tau'_j}^{ \b'_j m'_j \sigma'_j} | \hat{\mathcal{R}}^{\dagger} |w_{\mathbf{R}+\boldsymbol\tau_j}^{ \b_j m_j \sigma_j}\rangle\,.
\end{align}
Write in matrix form
\begin{align}
\langle \mathbf{0} |\hat{\mathcal{R}} \hat{H} \hat{\mathcal{R}}^{\dagger} | \mathbf{R} \rangle=\sum_{\mathbf{R}' \mathbf{R}^{\prime\prime}} \mathcal{R}_{\mathbf{0}\mathbf{R}'} H^{\mathbf{R}^{\prime\prime}-\mathbf{R}^{\prime}}\mathcal{R}_{\mathbf{R}^{\prime\prime}\mathbf{R}}\,.
\end{align}

If a Hamiltonian does not satisfy the system's symmetry, its symmetrized Hamiltonian can be obtained as the average over all symmetry operations given by
\begin{eqnarray}
  H^{sym}(\mathbf{R})=\dfrac{1}{|\mathcal{G}_H|}\sum_{\mathcal{R}\in \mathcal{G}_H} \langle \mathbf{0} |\hat{\mathcal{R}} \hat{H} \hat{\mathcal{R}}^{\dagger} | \mathbf{R} \rangle\,.
\end{eqnarray}
where $|\mathcal{G}_H|$ is the number of symmetry operators in the group $\mathcal{G}_H$.

\section{Calculate $J_{ij}$ by mapping Total energy calculations}\label{Jij_fittingE}
\label{jijtote}
For $\srmnsb$ in a $J_1$-$J_2$-$J_a$ model, to extract the three exchange parameters by energy-mapping method, we can calculate the total energies of four AFM magnetic configurations: C-, A-, G-, and Stripe(S)-type.
A supercell with basis vectors $\bold a$, $\bold b + \bold c$ and $\bold b - \bold c$ is constructed and used to simulate these four magnetic configurations.
According to the Heisenberg model as stated in \req{eq:h-jn}, we obtain
\begin{align}
  4E_\text{A}  /S^2 &=&  2J_{1} + &2J_{2} - J_{c}/2\,, \nn
  4E_\text{G}  /S^2 &=& -2J_{1} + &2J_{2} - J_{c}/2\,, \nn 
  4E_\text{S}  /S^2 &=&        - &2J_{2} - J_{c}/2\,, \nn
  4E_\text{C}  /S^2 &=& -2J_{1} + &2J_{2} + J_{c}/2\,,  
\end{align}
where the energies $E_\text{A}$, $E_\text{G}$, $E_\text{SA}$ and $E_\text{C}$ are the calculated total energy per magnetic atom for the four configurations. 
Here we do not distinguish $J_{2b}$ and $J_{2c}$, instead, an average of them as $J_{2}$ is calculated.
Then the exchange coupling constants are given by

\begin{align}
  J_\text{c}&=& 4(E_\text{C}-E_\text{G})/S^2\,,\nn
  J_1&=& (E_\text{A}-E_\text{G})/S^2\,,\nn
  J_2&=& (E_\text{A}-E_\text{S})/S^2 - J_1/2\,.
\end{align}

\section{Orbital resolved exchange couplings}
\begin{figure}[ht]
  \centering
  \begin{tabular}{c}
    \includegraphics[width=.9\linewidth,clip]{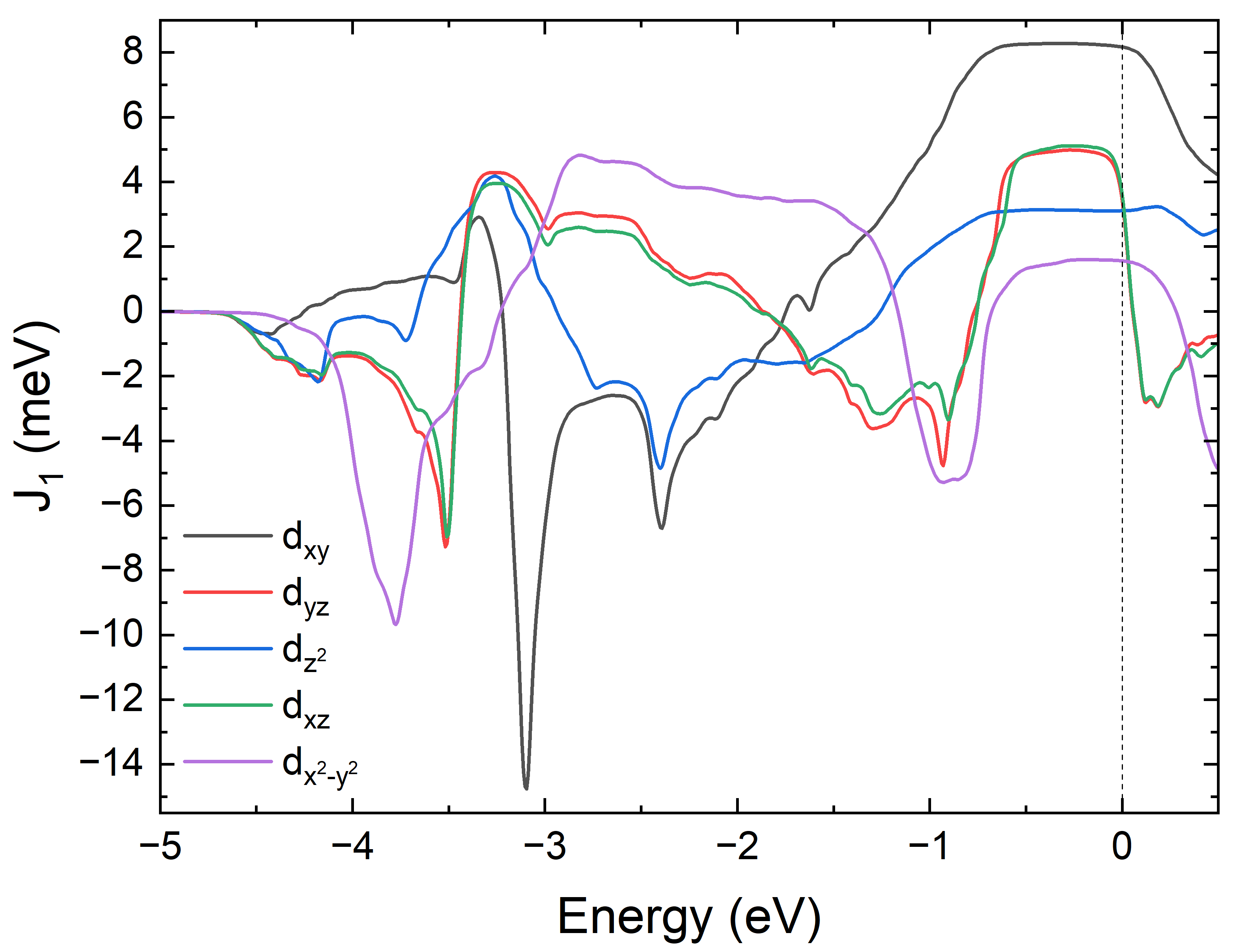} \\  
    \includegraphics[width=.9\linewidth,clip]{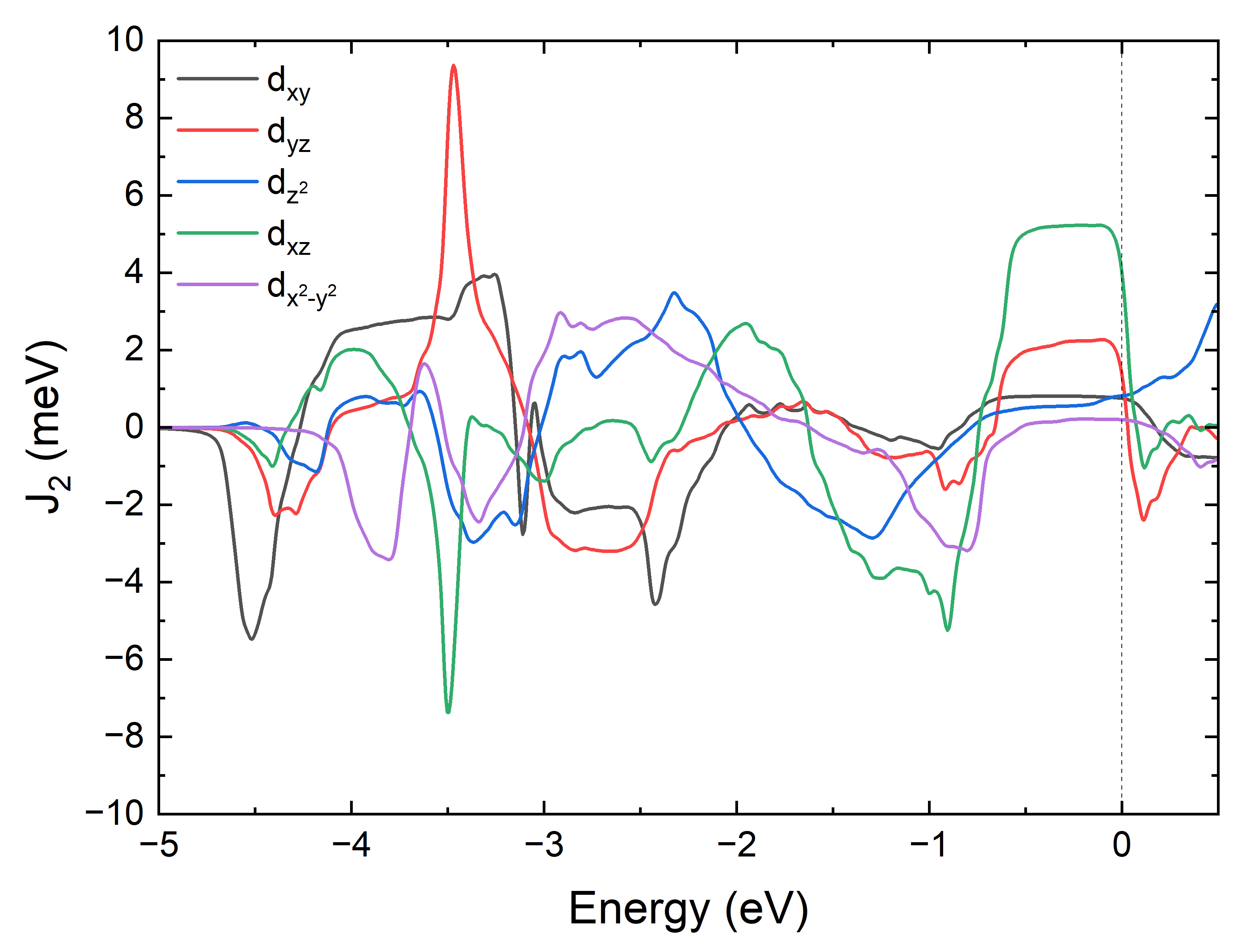} \\
  \end{tabular}%
  \caption{ Orbital-resolved exchange couplings of $\srmnsb$ as a function Fermi energy are plotted with color
    weights, with black identifying the Mn-$d_{xy}$ state, blue the Mn-$d_{3z^2-1}$ states, red the
    Mn-$d_{yz}$ states, green the Mn-$d_{xz}$ state and purple the Mn-$d_{x^2-y^2}$. The horizontal zero is the Fermi level.}
  \label{fig:j12cw_orb}
\end{figure}

%\bibliography{aaa.bib}

\begin{thebibliography}{46}%
\makeatletter
\providecommand \@ifxundefined [1]{%
 \@ifx{#1\undefined}
}%
\providecommand \@ifnum [1]{%
 \ifnum #1\expandafter \@firstoftwo
 \else \expandafter \@secondoftwo
 \fi
}%
\providecommand \@ifx [1]{%
 \ifx #1\expandafter \@firstoftwo
 \else \expandafter \@secondoftwo
 \fi
}%
\providecommand \natexlab [1]{#1}%
\providecommand \enquote  [1]{``#1''}%
\providecommand \bibnamefont  [1]{#1}%
\providecommand \bibfnamefont [1]{#1}%
\providecommand \citenamefont [1]{#1}%
\providecommand \href@noop [0]{\@secondoftwo}%
\providecommand \href [0]{\begingroup \@sanitize@url \@href}%
\providecommand \@href[1]{\@@startlink{#1}\@@href}%
\providecommand \@@href[1]{\endgroup#1\@@endlink}%
\providecommand \@sanitize@url [0]{\catcode `\\12\catcode `\$12\catcode
  `\&12\catcode `\#12\catcode `\^12\catcode `\_12\catcode `\%12\relax}%
\providecommand \@@startlink[1]{}%
\providecommand \@@endlink[0]{}%
\providecommand \url  [0]{\begingroup\@sanitize@url \@url }%
\providecommand \@url [1]{\endgroup\@href {#1}{\urlprefix }}%
\providecommand \urlprefix  [0]{URL }%
\providecommand \Eprint [0]{\href }%
\providecommand \doibase [0]{https://doi.org/}%
\providecommand \selectlanguage [0]{\@gobble}%
\providecommand \bibinfo  [0]{\@secondoftwo}%
\providecommand \bibfield  [0]{\@secondoftwo}%
\providecommand \translation [1]{[#1]}%
\providecommand \BibitemOpen [0]{}%
\providecommand \bibitemStop [0]{}%
\providecommand \bibitemNoStop [0]{.\EOS\space}%
\providecommand \EOS [0]{\spacefactor3000\relax}%
\providecommand \BibitemShut  [1]{\csname bibitem#1\endcsname}%
\let\auto@bib@innerbib\@empty
%</preamble>
\bibitem [{\citenamefont {Wilde}\ \emph {et~al.}(2022)\citenamefont {Wilde},
  \citenamefont {Riberolles}, \citenamefont {Das}, \citenamefont {Liu},
  \citenamefont {Heitmann}, \citenamefont {Wang}, \citenamefont {Straszheim},
  \citenamefont {Bud'ko}, \citenamefont {Canfield}, \citenamefont {Kreyssig},
  \citenamefont {McQueeney}, \citenamefont {Ryan},\ and\ \citenamefont
  {Ueland}}]{Wilde2022prb}%
  \BibitemOpen
  \bibfield  {author} {\bibinfo {author} {\bibfnamefont {J.~M.}\ \bibnamefont
  {Wilde}}, \bibinfo {author} {\bibfnamefont {S.~X.~M.}\ \bibnamefont
  {Riberolles}}, \bibinfo {author} {\bibfnamefont {A.}~\bibnamefont {Das}},
  \bibinfo {author} {\bibfnamefont {Y.}~\bibnamefont {Liu}}, \bibinfo {author}
  {\bibfnamefont {T.~W.}\ \bibnamefont {Heitmann}}, \bibinfo {author}
  {\bibfnamefont {X.}~\bibnamefont {Wang}}, \bibinfo {author} {\bibfnamefont
  {W.~E.}\ \bibnamefont {Straszheim}}, \bibinfo {author} {\bibfnamefont
  {S.~L.}\ \bibnamefont {Bud'ko}}, \bibinfo {author} {\bibfnamefont {P.~C.}\
  \bibnamefont {Canfield}}, \bibinfo {author} {\bibfnamefont {A.}~\bibnamefont
  {Kreyssig}}, \bibinfo {author} {\bibfnamefont {R.~J.}\ \bibnamefont
  {McQueeney}}, \bibinfo {author} {\bibfnamefont {D.~H.}\ \bibnamefont
  {Ryan}},\ and\ \bibinfo {author} {\bibfnamefont {B.~G.}\ \bibnamefont
  {Ueland}},\ }\bibfield  {title} {\bibinfo {title} {Canted antiferromagnetic
  phases in the candidate layered weyl material {EuMnSb}$_{2}$},\ }\href
  {https://doi.org/10.1103/PhysRevB.106.024420} {\bibfield  {journal} {\bibinfo
   {journal} {Phys. Rev. B}\ }\textbf {\bibinfo {volume} {106}},\ \bibinfo
  {pages} {024420} (\bibinfo {year} {2022})}\BibitemShut {NoStop}%
\bibitem [{\citenamefont {Hasan}\ and\ \citenamefont
  {Kane}(2010)}]{Hasan2010rmp}%
  \BibitemOpen
  \bibfield  {author} {\bibinfo {author} {\bibfnamefont {M.~Z.}\ \bibnamefont
  {Hasan}}\ and\ \bibinfo {author} {\bibfnamefont {C.~L.}\ \bibnamefont
  {Kane}},\ }\bibfield  {title} {\bibinfo {title} {Colloquium: Topological
  insulators},\ }\href {https://doi.org/10.1103/RevModPhys.82.3045} {\bibfield
  {journal} {\bibinfo  {journal} {Rev. Mod. Phys.}\ }\textbf {\bibinfo {volume}
  {82}},\ \bibinfo {pages} {3045} (\bibinfo {year} {2010})}\BibitemShut
  {NoStop}%
\bibitem [{\citenamefont {Bansil}\ \emph {et~al.}(2016)\citenamefont {Bansil},
  \citenamefont {Lin},\ and\ \citenamefont {Das}}]{Bansil2016rmp}%
  \BibitemOpen
  \bibfield  {author} {\bibinfo {author} {\bibfnamefont {A.}~\bibnamefont
  {Bansil}}, \bibinfo {author} {\bibfnamefont {H.}~\bibnamefont {Lin}},\ and\
  \bibinfo {author} {\bibfnamefont {T.}~\bibnamefont {Das}},\ }\bibfield
  {title} {\bibinfo {title} {Colloquium: Topological band theory},\ }\href
  {https://doi.org/10.1103/RevModPhys.88.021004} {\bibfield  {journal}
  {\bibinfo  {journal} {Rev. Mod. Phys.}\ }\textbf {\bibinfo {volume} {88}},\
  \bibinfo {pages} {021004} (\bibinfo {year} {2016})}\BibitemShut {NoStop}%
\bibitem [{\citenamefont {Armitage}\ \emph {et~al.}(2018)\citenamefont
  {Armitage}, \citenamefont {Mele},\ and\ \citenamefont
  {Vishwanath}}]{Armitage2018rmp}%
  \BibitemOpen
  \bibfield  {author} {\bibinfo {author} {\bibfnamefont {N.~P.}\ \bibnamefont
  {Armitage}}, \bibinfo {author} {\bibfnamefont {E.~J.}\ \bibnamefont {Mele}},\
  and\ \bibinfo {author} {\bibfnamefont {A.}~\bibnamefont {Vishwanath}},\
  }\bibfield  {title} {\bibinfo {title} {Weyl and dirac semimetals in
  three-dimensional solids},\ }\href
  {https://doi.org/10.1103/RevModPhys.90.015001} {\bibfield  {journal}
  {\bibinfo  {journal} {Rev. Mod. Phys.}\ }\textbf {\bibinfo {volume} {90}},\
  \bibinfo {pages} {015001} (\bibinfo {year} {2018})}\BibitemShut {NoStop}%
\bibitem [{\citenamefont {Lv}\ \emph {et~al.}(2021)\citenamefont {Lv},
  \citenamefont {Qian},\ and\ \citenamefont {Ding}}]{Lv2021rmp}%
  \BibitemOpen
  \bibfield  {author} {\bibinfo {author} {\bibfnamefont {B.~Q.}\ \bibnamefont
  {Lv}}, \bibinfo {author} {\bibfnamefont {T.}~\bibnamefont {Qian}},\ and\
  \bibinfo {author} {\bibfnamefont {H.}~\bibnamefont {Ding}},\ }\bibfield
  {title} {\bibinfo {title} {Experimental perspective on three-dimensional
  topological semimetals},\ }\href
  {https://doi.org/10.1103/RevModPhys.93.025002} {\bibfield  {journal}
  {\bibinfo  {journal} {Rev. Mod. Phys.}\ }\textbf {\bibinfo {volume} {93}},\
  \bibinfo {pages} {025002} (\bibinfo {year} {2021})}\BibitemShut {NoStop}%
\bibitem [{\citenamefont {Vanderbilt}(2018)}]{Vanderbilt2018}%
  \BibitemOpen
  \bibfield  {author} {\bibinfo {author} {\bibfnamefont {D.}~\bibnamefont
  {Vanderbilt}},\ }\href@noop {} {\emph {\bibinfo {title} {{Berry Phases in
  Electronic Structure Theory: Electric Polarization, Orbital Magnetization and
  Topological Insulators}}}}\ (\bibinfo  {publisher} {Cambridge University
  Press},\ \bibinfo {address} {Cambridge, England},\ \bibinfo {year}
  {2018})\BibitemShut {NoStop}%
\bibitem [{\citenamefont {Zhang}\ \emph {et~al.}(2019)\citenamefont {Zhang},
  \citenamefont {Okamoto}, \citenamefont {Stone}, \citenamefont {Liu},
  \citenamefont {Zhu}, \citenamefont {DiTusa}, \citenamefont {Mao},\ and\
  \citenamefont {Tennant}}]{zhang2019prb}%
  \BibitemOpen
  \bibfield  {author} {\bibinfo {author} {\bibfnamefont {Q.}~\bibnamefont
  {Zhang}}, \bibinfo {author} {\bibfnamefont {S.}~\bibnamefont {Okamoto}},
  \bibinfo {author} {\bibfnamefont {M.~B.}\ \bibnamefont {Stone}}, \bibinfo
  {author} {\bibfnamefont {J.}~\bibnamefont {Liu}}, \bibinfo {author}
  {\bibfnamefont {Y.}~\bibnamefont {Zhu}}, \bibinfo {author} {\bibfnamefont
  {J.}~\bibnamefont {DiTusa}}, \bibinfo {author} {\bibfnamefont
  {Z.}~\bibnamefont {Mao}},\ and\ \bibinfo {author} {\bibfnamefont {D.~A.}\
  \bibnamefont {Tennant}},\ }\bibfield  {title} {\bibinfo {title} {Influence of
  magnetism on dirac semimetallic behavior in nonstoichiometric
  {Sr}$_{1\ensuremath{-}y}${Mn}$_{1\ensuremath{-}z}${Sb}$_{2}\phantom{\rule{4pt}{0ex}}(y\ensuremath{\sim}0.07,z\ensuremath{\sim}0.02)$},\
  }\href {https://doi.org/10.1103/physrevb.100.205105} {\bibfield  {journal}
  {\bibinfo  {journal} {Phys. Rev. B}\ }\textbf {\bibinfo {volume} {100}},\
  \bibinfo {pages} {205105} (\bibinfo {year} {2019})}\BibitemShut {NoStop}%
\bibitem [{\citenamefont {Ceccatto}\ \emph {et~al.}(1992)\citenamefont
  {Ceccatto}, \citenamefont {Gazza},\ and\ \citenamefont
  {Trumper}}]{ceccatto1992prb}%
  \BibitemOpen
  \bibfield  {author} {\bibinfo {author} {\bibfnamefont {H.~A.}\ \bibnamefont
  {Ceccatto}}, \bibinfo {author} {\bibfnamefont {C.~J.}\ \bibnamefont
  {Gazza}},\ and\ \bibinfo {author} {\bibfnamefont {A.~E.}\ \bibnamefont
  {Trumper}},\ }\bibfield  {title} {\bibinfo {title} {{$J_1$-$J_2$ model:
  Energy, correlations, and order-parameter fluctuations on finite lattices}},\
  }\href {https://doi.org/10.1103/PhysRevB.45.7832} {\bibfield  {journal}
  {\bibinfo  {journal} {Phys. Rev. B}\ }\textbf {\bibinfo {volume} {45}},\
  \bibinfo {pages} {7832} (\bibinfo {year} {1992})}\BibitemShut {NoStop}%
\bibitem [{\citenamefont {Sirker}\ \emph {et~al.}(2006)\citenamefont {Sirker},
  \citenamefont {Weihong}, \citenamefont {Sushkov},\ and\ \citenamefont
  {Oitmaa}}]{sirker2006prb}%
  \BibitemOpen
  \bibfield  {author} {\bibinfo {author} {\bibfnamefont {J.}~\bibnamefont
  {Sirker}}, \bibinfo {author} {\bibfnamefont {Z.}~\bibnamefont {Weihong}},
  \bibinfo {author} {\bibfnamefont {O.~P.}\ \bibnamefont {Sushkov}},\ and\
  \bibinfo {author} {\bibfnamefont {J.}~\bibnamefont {Oitmaa}},\ }\bibfield
  {title} {\bibinfo {title} {{$J1$-$J2$ model: First-order phase transition
  versus deconfinement of spinons}},\ }\href
  {https://doi.org/10.1103/physrevb.73.184420} {\bibfield  {journal} {\bibinfo
  {journal} {Physical Review B}\ }\textbf {\bibinfo {volume} {73}},\ \bibinfo
  {pages} {184420} (\bibinfo {year} {2006})}\BibitemShut {NoStop}%
\bibitem [{\citenamefont {Liechtenstein}\ \emph {et~al.}(1987)\citenamefont
  {Liechtenstein}, \citenamefont {Katsnelson}, \citenamefont {Antropov},\ and\
  \citenamefont {Gubanov}}]{liechtenstein1987jmmm}%
  \BibitemOpen
  \bibfield  {author} {\bibinfo {author} {\bibfnamefont {A.}~\bibnamefont
  {Liechtenstein}}, \bibinfo {author} {\bibfnamefont {M.}~\bibnamefont
  {Katsnelson}}, \bibinfo {author} {\bibfnamefont {V.}~\bibnamefont
  {Antropov}},\ and\ \bibinfo {author} {\bibfnamefont {V.}~\bibnamefont
  {Gubanov}},\ }\bibfield  {title} {\bibinfo {title} {Local spin density
  functional approach to the theory of exchange interactions in ferromagnetic
  metals and alloys},\ }\href
  {https://doi.org/https://doi.org/10.1016/0304-8853(87)90721-9} {\bibfield
  {journal} {\bibinfo  {journal} {Journal of Magnetism and Magnetic Materials}\
  }\textbf {\bibinfo {volume} {67}},\ \bibinfo {pages} {65 } (\bibinfo {year}
  {1987})}\BibitemShut {NoStop}%
\bibitem [{\citenamefont {Kotani}\ and\ \citenamefont {van
  Schilfgaarde}(2008)}]{kotani2008jpcm}%
  \BibitemOpen
  \bibfield  {author} {\bibinfo {author} {\bibfnamefont {T.}~\bibnamefont
  {Kotani}}\ and\ \bibinfo {author} {\bibfnamefont {M.}~\bibnamefont {van
  Schilfgaarde}},\ }\bibfield  {title} {\bibinfo {title} {Spin wave dispersion
  based on the quasiparticle self-consistent gw method: {NiO}, {MnO} and
  $\alpha$-{MnAs}},\ }\href {https://doi.org/10.1088/0953-8984/20/29/295214}
  {\bibfield  {journal} {\bibinfo  {journal} {Journal of Physics: Condensed
  Matter}\ }\textbf {\bibinfo {volume} {20}},\ \bibinfo {pages} {295214}
  (\bibinfo {year} {2008})}\BibitemShut {NoStop}%
\bibitem [{\citenamefont {Ye}\ \emph {et~al.}(2021)\citenamefont {Ye},
  \citenamefont {Morgan}, \citenamefont {Tian}, \citenamefont {Chi},
  \citenamefont {Wang}, \citenamefont {Manley}, \citenamefont {Parker},
  \citenamefont {Khan}, \citenamefont {Mitchell},\ and\ \citenamefont
  {Fishman}}]{ye2021prb}%
  \BibitemOpen
  \bibfield  {author} {\bibinfo {author} {\bibfnamefont {F.}~\bibnamefont
  {Ye}}, \bibinfo {author} {\bibfnamefont {Z.}~\bibnamefont {Morgan}}, \bibinfo
  {author} {\bibfnamefont {W.}~\bibnamefont {Tian}}, \bibinfo {author}
  {\bibfnamefont {S.}~\bibnamefont {Chi}}, \bibinfo {author} {\bibfnamefont
  {X.}~\bibnamefont {Wang}}, \bibinfo {author} {\bibfnamefont {M.~E.}\
  \bibnamefont {Manley}}, \bibinfo {author} {\bibfnamefont {D.}~\bibnamefont
  {Parker}}, \bibinfo {author} {\bibfnamefont {M.~A.}\ \bibnamefont {Khan}},
  \bibinfo {author} {\bibfnamefont {J.~F.}\ \bibnamefont {Mitchell}},\ and\
  \bibinfo {author} {\bibfnamefont {R.}~\bibnamefont {Fishman}},\ }\bibfield
  {title} {\bibinfo {title} {Canted antiferromagnetic order and spin dynamics
  in the honeycomb-lattice compound {Tb}$_{2}${Ir}$_{3}${Ga}$_{9}$},\ }\href
  {https://doi.org/10.1103/PhysRevB.103.184413} {\bibfield  {journal} {\bibinfo
   {journal} {Phys. Rev. B}\ }\textbf {\bibinfo {volume} {103}},\ \bibinfo
  {pages} {184413} (\bibinfo {year} {2021})}\BibitemShut {NoStop}%
\bibitem [{\citenamefont {{van Schilfgaarde}}\ \emph
  {et~al.}(2006)\citenamefont {{van Schilfgaarde}}, \citenamefont {Kotani},\
  and\ \citenamefont {Faleev}}]{van-schilfgaarde2006prl}%
  \BibitemOpen
  \bibfield  {author} {\bibinfo {author} {\bibfnamefont {M.}~\bibnamefont {{van
  Schilfgaarde}}}, \bibinfo {author} {\bibfnamefont {T.}~\bibnamefont
  {Kotani}},\ and\ \bibinfo {author} {\bibfnamefont {S.}~\bibnamefont
  {Faleev}},\ }\bibfield  {title} {\bibinfo {title} {Quasiparticle
  self-consistent {GW} theory},\ }\href
  {https://doi.org/10.1103/PhysRevLett.96.226402} {\bibfield  {journal}
  {\bibinfo  {journal} {Phys. Rev. Lett.}\ }\textbf {\bibinfo {volume} {96}},\
  \bibinfo {pages} {226402} (\bibinfo {year} {2006})}\BibitemShut {NoStop}%
\bibitem [{\citenamefont {Ke}\ \emph {et~al.}(2011)\citenamefont {Ke},
  \citenamefont {van Schilfgaarde}, \citenamefont {Pulikkotil}, \citenamefont
  {Kotani},\ and\ \citenamefont {Antropov}}]{ke2011prbr}%
  \BibitemOpen
  \bibfield  {author} {\bibinfo {author} {\bibfnamefont {L.}~\bibnamefont
  {Ke}}, \bibinfo {author} {\bibfnamefont {M.}~\bibnamefont {van
  Schilfgaarde}}, \bibinfo {author} {\bibfnamefont {J.}~\bibnamefont
  {Pulikkotil}}, \bibinfo {author} {\bibfnamefont {T.}~\bibnamefont {Kotani}},\
  and\ \bibinfo {author} {\bibfnamefont {V.}~\bibnamefont {Antropov}},\
  }\bibfield  {title} {\bibinfo {title} {Low-energy coherent stoner-like
  excitations in {CaFe$_{2}$As$_{2}$}},\ }\href
  {https://doi.org/10.1103/PhysRevB.83.060404} {\bibfield  {journal} {\bibinfo
  {journal} {Phys. Rev. B: Rapid communication}\ }\textbf {\bibinfo {volume}
  {83}},\ \bibinfo {pages} {060404(R)} (\bibinfo {year} {2011})}\BibitemShut
  {NoStop}%
\bibitem [{\citenamefont {Ke}\ and\ \citenamefont
  {Katsnelson}(2021)}]{ke2021ncm}%
  \BibitemOpen
  \bibfield  {author} {\bibinfo {author} {\bibfnamefont {L.}~\bibnamefont
  {Ke}}\ and\ \bibinfo {author} {\bibfnamefont {M.~I.}\ \bibnamefont
  {Katsnelson}},\ }\bibfield  {title} {\bibinfo {title} {{Electron correlation
  effects on exchange interactions and spin excitations in 2D van der Waals
  materials}},\ }\href {https://doi.org/10.1038/s41524-020-00469-2} {\bibfield
  {journal} {\bibinfo  {journal} {npj Computational Materials}\ }\textbf
  {\bibinfo {volume} {7}},\ \bibinfo {pages} {1} (\bibinfo {year}
  {2021})}\BibitemShut {NoStop}%
\bibitem [{\citenamefont {Marzari}\ and\ \citenamefont
  {Vanderbilt}(1997)}]{marzari1997prb}%
  \BibitemOpen
  \bibfield  {author} {\bibinfo {author} {\bibfnamefont {N.}~\bibnamefont
  {Marzari}}\ and\ \bibinfo {author} {\bibfnamefont {D.}~\bibnamefont
  {Vanderbilt}},\ }\bibfield  {title} {\bibinfo {title} {Maximally localized
  generalized wannier functions for composite energy bands},\ }\href
  {https://doi.org/10.1103/physrevb.56.12847} {\bibfield  {journal} {\bibinfo
  {journal} {Physical Review B}\ }\textbf {\bibinfo {volume} {56}},\ \bibinfo
  {pages} {12847} (\bibinfo {year} {1997})}\BibitemShut {NoStop}%
\bibitem [{\citenamefont {Souza}\ \emph {et~al.}(2001)\citenamefont {Souza},
  \citenamefont {Marzari},\ and\ \citenamefont {Vanderbilt}}]{souza2001prb}%
  \BibitemOpen
  \bibfield  {author} {\bibinfo {author} {\bibfnamefont {I.}~\bibnamefont
  {Souza}}, \bibinfo {author} {\bibfnamefont {N.}~\bibnamefont {Marzari}},\
  and\ \bibinfo {author} {\bibfnamefont {D.}~\bibnamefont {Vanderbilt}},\
  }\bibfield  {title} {\bibinfo {title} {Maximally localized wannier functions
  for entangled energy bands},\ }\href
  {https://doi.org/10.1103/PhysRevB.65.035109} {\bibfield  {journal} {\bibinfo
  {journal} {Phys. Rev. B}\ }\textbf {\bibinfo {volume} {65}},\ \bibinfo
  {pages} {035109} (\bibinfo {year} {2001})}\BibitemShut {NoStop}%
\bibitem [{\citenamefont {Marzari}\ \emph {et~al.}(2012)\citenamefont
  {Marzari}, \citenamefont {Mostofi}, \citenamefont {Yates}, \citenamefont
  {Souza},\ and\ \citenamefont {Vanderbilt}}]{marzari2012rmp}%
  \BibitemOpen
  \bibfield  {author} {\bibinfo {author} {\bibfnamefont {N.}~\bibnamefont
  {Marzari}}, \bibinfo {author} {\bibfnamefont {A.~A.}\ \bibnamefont
  {Mostofi}}, \bibinfo {author} {\bibfnamefont {J.~R.}\ \bibnamefont {Yates}},
  \bibinfo {author} {\bibfnamefont {I.}~\bibnamefont {Souza}},\ and\ \bibinfo
  {author} {\bibfnamefont {D.}~\bibnamefont {Vanderbilt}},\ }\bibfield  {title}
  {\bibinfo {title} {Maximally localized wannier functions: Theory and
  applications},\ }\href {https://doi.org/10.1103/RevModPhys.84.1419}
  {\bibfield  {journal} {\bibinfo  {journal} {Rev. Mod. Phys.}\ }\textbf
  {\bibinfo {volume} {84}},\ \bibinfo {pages} {1419} (\bibinfo {year}
  {2012})}\BibitemShut {NoStop}%
\bibitem [{\citenamefont {Ke}(2019)}]{ke2019prb}%
  \BibitemOpen
  \bibfield  {author} {\bibinfo {author} {\bibfnamefont {L.}~\bibnamefont
  {Ke}},\ }\bibfield  {title} {\bibinfo {title} {Intersublattice
  magnetocrystalline anisotropy using a realistic tight-binding method based on
  maximally localized {Wannier} functions},\ }\href
  {https://doi.org/10.1103/PhysRevB.99.054418} {\bibfield  {journal} {\bibinfo
  {journal} {Phys. Rev. B}\ }\textbf {\bibinfo {volume} {99}},\ \bibinfo
  {pages} {054418} (\bibinfo {year} {2019})}\BibitemShut {NoStop}%
\bibitem [{\citenamefont {Brechtel}\ \emph {et~al.}(1981)\citenamefont
  {Brechtel}, \citenamefont {Cordier},\ and\ \citenamefont
  {Sch\"{a}fer}}]{brechtel1981jlcm}%
  \BibitemOpen
  \bibfield  {author} {\bibinfo {author} {\bibfnamefont {E.}~\bibnamefont
  {Brechtel}}, \bibinfo {author} {\bibfnamefont {G.}~\bibnamefont {Cordier}},\
  and\ \bibinfo {author} {\bibfnamefont {H.}~\bibnamefont {Sch\"{a}fer}},\
  }\bibfield  {title} {\bibinfo {title} {Neue tern\"{a}re
  erdalkali-übergangselement-pnictide},\ }\href
  {https://doi.org/10.1016/0022-5088(81)90057-6} {\bibfield  {journal}
  {\bibinfo  {journal} {Journal of the Less Common Metals}\ }\textbf {\bibinfo
  {volume} {79}},\ \bibinfo {pages} {131} (\bibinfo {year} {1981})}\BibitemShut
  {NoStop}%
\bibitem [{\citenamefont {Ramankutty}\ \emph {et~al.}(2018)\citenamefont
  {Ramankutty}, \citenamefont {Henke}, \citenamefont {Schiphorst},
  \citenamefont {Nutakki}, \citenamefont {Bron}, \citenamefont
  {Araizi-Kanoutas}, \citenamefont {Mishra}, \citenamefont {Li}, \citenamefont
  {Huang}, \citenamefont {Kim}, \citenamefont {Hoesch}, \citenamefont
  {Schlueter}, \citenamefont {Lee}, \citenamefont {de~Visser}, \citenamefont
  {Zhong}, \citenamefont {van Wezel}, \citenamefont {van Heumen},\ and\
  \citenamefont {Golden}}]{ramankutty2018sp}%
  \BibitemOpen
  \bibfield  {author} {\bibinfo {author} {\bibfnamefont {S.~V.}\ \bibnamefont
  {Ramankutty}}, \bibinfo {author} {\bibfnamefont {J.}~\bibnamefont {Henke}},
  \bibinfo {author} {\bibfnamefont {A.}~\bibnamefont {Schiphorst}}, \bibinfo
  {author} {\bibfnamefont {R.}~\bibnamefont {Nutakki}}, \bibinfo {author}
  {\bibfnamefont {S.}~\bibnamefont {Bron}}, \bibinfo {author} {\bibfnamefont
  {G.}~\bibnamefont {Araizi-Kanoutas}}, \bibinfo {author} {\bibfnamefont
  {S.}~\bibnamefont {Mishra}}, \bibinfo {author} {\bibfnamefont
  {L.}~\bibnamefont {Li}}, \bibinfo {author} {\bibfnamefont {Y.}~\bibnamefont
  {Huang}}, \bibinfo {author} {\bibfnamefont {T.}~\bibnamefont {Kim}}, \bibinfo
  {author} {\bibfnamefont {M.}~\bibnamefont {Hoesch}}, \bibinfo {author}
  {\bibfnamefont {C.}~\bibnamefont {Schlueter}}, \bibinfo {author}
  {\bibfnamefont {T.-L.}\ \bibnamefont {Lee}}, \bibinfo {author} {\bibfnamefont
  {A.}~\bibnamefont {de~Visser}}, \bibinfo {author} {\bibfnamefont
  {Z.}~\bibnamefont {Zhong}}, \bibinfo {author} {\bibfnamefont
  {J.}~\bibnamefont {van Wezel}}, \bibinfo {author} {\bibfnamefont
  {E.}~\bibnamefont {van Heumen}},\ and\ \bibinfo {author} {\bibfnamefont
  {M.}~\bibnamefont {Golden}},\ }\bibfield  {title} {\bibinfo {title}
  {{Electronic structure of the candidate 2D Dirac semimetal SrMnSb$_2$: a
  combined experimental and theoretical study}},\ }\bibfield  {journal}
  {\bibinfo  {journal} {{SciPost} Physics}\ }\textbf {\bibinfo {volume} {4}},\
  \href {https://doi.org/10.21468/scipostphys.4.2.010}
  {10.21468/scipostphys.4.2.010} (\bibinfo {year} {2018})\BibitemShut {NoStop}%
\bibitem [{\citenamefont {Li}\ \emph {et~al.}(2021)\citenamefont {Li},
  \citenamefont {Pajerowski}, \citenamefont {Riberolles}, \citenamefont {Ke},
  \citenamefont {Yan},\ and\ \citenamefont {McQueeney}}]{li2021prb}%
  \BibitemOpen
  \bibfield  {author} {\bibinfo {author} {\bibfnamefont {B.}~\bibnamefont
  {Li}}, \bibinfo {author} {\bibfnamefont {D.~M.}\ \bibnamefont {Pajerowski}},
  \bibinfo {author} {\bibfnamefont {S.~X.~M.}\ \bibnamefont {Riberolles}},
  \bibinfo {author} {\bibfnamefont {L.}~\bibnamefont {Ke}}, \bibinfo {author}
  {\bibfnamefont {J.-Q.}\ \bibnamefont {Yan}},\ and\ \bibinfo {author}
  {\bibfnamefont {R.~J.}\ \bibnamefont {McQueeney}},\ }\bibfield  {title}
  {\bibinfo {title} {Quasi-two-dimensional ferromagnetism and anisotropic
  interlayer couplings in the magnetic topological insulator
  ${\mathrm{mnbi}}_{2}{\mathrm{te}}_{4}$},\ }\href
  {https://doi.org/10.1103/PhysRevB.104.L220402} {\bibfield  {journal}
  {\bibinfo  {journal} {Phys. Rev. B}\ }\textbf {\bibinfo {volume} {104}},\
  \bibinfo {pages} {L220402} (\bibinfo {year} {2021})}\BibitemShut {NoStop}%
\bibitem [{\citenamefont {Mkhitaryan}\ and\ \citenamefont
  {Ke}(2021)}]{mkhitaryan2021prb}%
  \BibitemOpen
  \bibfield  {author} {\bibinfo {author} {\bibfnamefont {V.~V.}\ \bibnamefont
  {Mkhitaryan}}\ and\ \bibinfo {author} {\bibfnamefont {L.}~\bibnamefont
  {Ke}},\ }\bibfield  {title} {\bibinfo {title} {Self-consistently renormalized
  spin-wave theory of layered ferromagnets on the honeycomb lattice},\ }\href
  {https://doi.org/10.1103/PhysRevB.104.064435} {\bibfield  {journal} {\bibinfo
   {journal} {Phys. Rev. B}\ }\textbf {\bibinfo {volume} {104}},\ \bibinfo
  {pages} {064435} (\bibinfo {year} {2021})}\BibitemShut {NoStop}%
\bibitem [{\citenamefont {Perdew}\ \emph {et~al.}(1996)\citenamefont {Perdew},
  \citenamefont {Burke},\ and\ \citenamefont {Ernzerhof}}]{perdew1996prl}%
  \BibitemOpen
  \bibfield  {author} {\bibinfo {author} {\bibfnamefont {J.~P.}\ \bibnamefont
  {Perdew}}, \bibinfo {author} {\bibfnamefont {K.}~\bibnamefont {Burke}},\ and\
  \bibinfo {author} {\bibfnamefont {M.}~\bibnamefont {Ernzerhof}},\ }\bibfield
  {title} {\bibinfo {title} {Generalized gradient approximation made simple},\
  }\href {https://doi.org/10.1103/PhysRevLett.77.3865} {\bibfield  {journal}
  {\bibinfo  {journal} {Phys. Rev. Lett.}\ }\textbf {\bibinfo {volume} {77}},\
  \bibinfo {pages} {3865} (\bibinfo {year} {1996})}\BibitemShut {NoStop}%
\bibitem [{\citenamefont {Kresse}\ and\ \citenamefont
  {Joubert}(1999)}]{kresse1999prb}%
  \BibitemOpen
  \bibfield  {author} {\bibinfo {author} {\bibfnamefont {G.}~\bibnamefont
  {Kresse}}\ and\ \bibinfo {author} {\bibfnamefont {D.}~\bibnamefont
  {Joubert}},\ }\bibfield  {title} {\bibinfo {title} {From ultrasoft
  pseudopotentials to the projector augmented-wave method},\ }\href
  {https://doi.org/10.1103/PhysRevB.59.1758} {\bibfield  {journal} {\bibinfo
  {journal} {Phys. Rev. B}\ }\textbf {\bibinfo {volume} {59}},\ \bibinfo
  {pages} {1758} (\bibinfo {year} {1999})}\BibitemShut {NoStop}%
\bibitem [{\citenamefont {Kresse}\ and\ \citenamefont
  {Furthm\"uller}(1996)}]{kresse1996prb}%
  \BibitemOpen
  \bibfield  {author} {\bibinfo {author} {\bibfnamefont {G.}~\bibnamefont
  {Kresse}}\ and\ \bibinfo {author} {\bibfnamefont {J.}~\bibnamefont
  {Furthm\"uller}},\ }\bibfield  {title} {\bibinfo {title} {Efficient iterative
  schemes for ab initio total-energy calculations using a plane-wave basis
  set},\ }\href {https://doi.org/10.1103/PhysRevB.54.11169} {\bibfield
  {journal} {\bibinfo  {journal} {Phys. Rev. B}\ }\textbf {\bibinfo {volume}
  {54}},\ \bibinfo {pages} {11169} (\bibinfo {year} {1996})}\BibitemShut
  {NoStop}%
\bibitem [{\citenamefont {Mostofi}\ \emph {et~al.}(2014)\citenamefont
  {Mostofi}, \citenamefont {Yates}, \citenamefont {Pizzi}, \citenamefont {Lee},
  \citenamefont {Souza}, \citenamefont {Vanderbilt},\ and\ \citenamefont
  {Marzari}}]{mostofi2014cpc}%
  \BibitemOpen
  \bibfield  {author} {\bibinfo {author} {\bibfnamefont {A.~A.}\ \bibnamefont
  {Mostofi}}, \bibinfo {author} {\bibfnamefont {J.~R.}\ \bibnamefont {Yates}},
  \bibinfo {author} {\bibfnamefont {G.}~\bibnamefont {Pizzi}}, \bibinfo
  {author} {\bibfnamefont {Y.-S.}\ \bibnamefont {Lee}}, \bibinfo {author}
  {\bibfnamefont {I.}~\bibnamefont {Souza}}, \bibinfo {author} {\bibfnamefont
  {D.}~\bibnamefont {Vanderbilt}},\ and\ \bibinfo {author} {\bibfnamefont
  {N.}~\bibnamefont {Marzari}},\ }\bibfield  {title} {\bibinfo {title} {An
  updated version of wannier90: A tool for obtaining maximally-localised
  wannier functions},\ }\href
  {https://doi.org/http://doi.org/10.1016/j.cpc.2014.05.003} {\bibfield
  {journal} {\bibinfo  {journal} {Computer Physics Communications}\ }\textbf
  {\bibinfo {volume} {185}},\ \bibinfo {pages} {2309} (\bibinfo {year}
  {2014})}\BibitemShut {NoStop}%
\bibitem [{\citenamefont {van Schilfgaarde}\ and\ \citenamefont
  {Antropov}(1999)}]{van-schilfgaarde1999jap}%
  \BibitemOpen
  \bibfield  {author} {\bibinfo {author} {\bibfnamefont {M.}~\bibnamefont {van
  Schilfgaarde}}\ and\ \bibinfo {author} {\bibfnamefont {V.~P.}\ \bibnamefont
  {Antropov}},\ }\bibfield  {title} {\bibinfo {title} {First-principles
  exchange interactions in {Fe}, {Ni}, and {Co}},\ }\href
  {https://doi.org/10.1063/1.370495} {\bibfield  {journal} {\bibinfo  {journal}
  {Journal of Applied Physics}\ }\textbf {\bibinfo {volume} {85}},\ \bibinfo
  {pages} {4827} (\bibinfo {year} {1999})}\BibitemShut {NoStop}%
\bibitem [{\citenamefont {Ke}\ and\ \citenamefont {van
  Schilfgaarde}(2012)}]{ke2012prbr}%
  \BibitemOpen
  \bibfield  {author} {\bibinfo {author} {\bibfnamefont {L.}~\bibnamefont
  {Ke}}\ and\ \bibinfo {author} {\bibfnamefont {Mark}~\bibnamefont {van
  Schilfgaarde}}\ and\ \bibinfo {author} {\bibfnamefont {Vladimir} ~\bibnamefont {Antropov}},\ }\bibfield  {title}
  {\bibinfo {title} {Spin excitations in {K$_2$Fe$_{4+x}$Se$_5$}: Linear
  response approach},\ }\href {https://doi.org/10.1103/PhysRevB.86.020402}
  {\bibfield  {journal} {\bibinfo  {journal} {Phys. Rev. B: Rapid
  communication}\ }\textbf {\bibinfo {volume} {86}},\ \bibinfo {pages} {020402(R)}
  (\bibinfo {year} {2012})}\BibitemShut {NoStop}%
\bibitem [{\citenamefont {Ke}\ \emph {et~al.}(2017)\citenamefont {Ke},
  \citenamefont {Harmon},\ and\ \citenamefont {Kramer}}]{ke2017prb}%
  \BibitemOpen
  \bibfield  {author} {\bibinfo {author} {\bibfnamefont {L.}~\bibnamefont
  {Ke}}, \bibinfo {author} {\bibfnamefont {B.~N.}\ \bibnamefont {Harmon}},\
  and\ \bibinfo {author} {\bibfnamefont {M.~J.}\ \bibnamefont {Kramer}},\
  }\bibfield  {title} {\bibinfo {title} {Electronic structure and magnetic
  properties in {${T}_{2}{\mathrm{AlB}}_{2}$ ($T$=Fe, Mn, Cr, Co, and Ni) and
  their alloys}},\ }\href {https://doi.org/10.1103/PhysRevB.95.104427}
  {\bibfield  {journal} {\bibinfo  {journal} {Phys. Rev. B}\ }\textbf {\bibinfo
  {volume} {95}},\ \bibinfo {pages} {104427} (\bibinfo {year}
  {2017})}\BibitemShut {NoStop}%
\bibitem [{\citenamefont {Ke}\ and\ \citenamefont {Johnson}(2016)}]{ke2016prb}%
  \BibitemOpen
  \bibfield  {author} {\bibinfo {author} {\bibfnamefont {L.}~\bibnamefont
  {Ke}}\ and\ \bibinfo {author} {\bibfnamefont {D.~D.}\ \bibnamefont
  {Johnson}},\ }\bibfield  {title} {\bibinfo {title} {Intrinsic magnetic
  properties in {$R$(Fe$_{1-x}$Co$_x$)$_{11}$Ti$Z$ ($R$=Y and Ce; $Z$=H, C, and
  N)}},\ }\href {https://doi.org/10.1103/PhysRevB.94.024423} {\bibfield
  {journal} {\bibinfo  {journal} {Phys. Rev. B}\ }\textbf {\bibinfo {volume}
  {94}},\ \bibinfo {pages} {024423} (\bibinfo {year} {2016})}\BibitemShut
  {NoStop}%
\bibitem [{\citenamefont {Ke}\ \emph {et~al.}(2013)\citenamefont {Ke},
  \citenamefont {Belashchenko}, \citenamefont {van Schilfgaarde}, \citenamefont
  {Kotani},\ and\ \citenamefont {Antropov}}]{ke2013prb}%
  \BibitemOpen
  \bibfield  {author} {\bibinfo {author} {\bibfnamefont {L.}~\bibnamefont
  {Ke}}, \bibinfo {author} {\bibfnamefont {K.~D.}\ \bibnamefont
  {Belashchenko}}, \bibinfo {author} {\bibfnamefont {M.}~\bibnamefont {van
  Schilfgaarde}}, \bibinfo {author} {\bibfnamefont {T.}~\bibnamefont
  {Kotani}},\ and\ \bibinfo {author} {\bibfnamefont {V.~P.}\ \bibnamefont
  {Antropov}},\ }\bibfield  {title} {\bibinfo {title} {Effects of alloying and
  strain on the magnetic properties of {Fe$_{16}$N$_2$}},\ }\href
  {https://doi.org/10.1103/PhysRevB.88.024404} {\bibfield  {journal} {\bibinfo
  {journal} {Phys. Rev. B}\ }\textbf {\bibinfo {volume} {88}},\ \bibinfo
  {pages} {024404} (\bibinfo {year} {2013})}\BibitemShut {NoStop}%
\bibitem [{\citenamefont {Pizzi}\ \emph {et~al.}(2020)\citenamefont {Pizzi},
  \citenamefont {Vitale}, \citenamefont {Arita}, \citenamefont {Bl\"{u}gel},
  \citenamefont {Freimuth}, \citenamefont {G{\'{e}}ranton}, \citenamefont
  {Gibertini}, \citenamefont {Gresch}, \citenamefont {Johnson}, \citenamefont
  {Koretsune}, \citenamefont {Iba{\~{n}}ez-Azpiroz}, \citenamefont {Lee},
  \citenamefont {Lihm}, \citenamefont {Marchand}, \citenamefont {Marrazzo},
  \citenamefont {Mokrousov}, \citenamefont {Mustafa}, \citenamefont {Nohara},
  \citenamefont {Nomura}, \citenamefont {Paulatto}, \citenamefont
  {Ponc{\'{e}}}, \citenamefont {Ponweiser}, \citenamefont {Qiao}, \citenamefont
  {Th\"{o}le}, \citenamefont {Tsirkin}, \citenamefont {Wierzbowska},
  \citenamefont {Marzari}, \citenamefont {Vanderbilt}, \citenamefont {Souza},
  \citenamefont {Mostofi},\ and\ \citenamefont {Yates}}]{pizzi2020jpcm}%
  \BibitemOpen
  \bibfield  {author} {\bibinfo {author} {\bibfnamefont {G.}~\bibnamefont
  {Pizzi}}, \bibinfo {author} {\bibfnamefont {V.}~\bibnamefont {Vitale}},
  \bibinfo {author} {\bibfnamefont {R.}~\bibnamefont {Arita}}, \bibinfo
  {author} {\bibfnamefont {S.}~\bibnamefont {Bl\"{u}gel}}, \bibinfo {author}
  {\bibfnamefont {F.}~\bibnamefont {Freimuth}}, \bibinfo {author}
  {\bibfnamefont {G.}~\bibnamefont {G{\'{e}}ranton}}, \bibinfo {author}
  {\bibfnamefont {M.}~\bibnamefont {Gibertini}}, \bibinfo {author}
  {\bibfnamefont {D.}~\bibnamefont {Gresch}}, \bibinfo {author} {\bibfnamefont
  {C.}~\bibnamefont {Johnson}}, \bibinfo {author} {\bibfnamefont
  {T.}~\bibnamefont {Koretsune}}, \bibinfo {author} {\bibfnamefont
  {J.}~\bibnamefont {Iba{\~{n}}ez-Azpiroz}}, \bibinfo {author} {\bibfnamefont
  {H.}~\bibnamefont {Lee}}, \bibinfo {author} {\bibfnamefont {J.-M.}\
  \bibnamefont {Lihm}}, \bibinfo {author} {\bibfnamefont {D.}~\bibnamefont
  {Marchand}}, \bibinfo {author} {\bibfnamefont {A.}~\bibnamefont {Marrazzo}},
  \bibinfo {author} {\bibfnamefont {Y.}~\bibnamefont {Mokrousov}}, \bibinfo
  {author} {\bibfnamefont {J.~I.}\ \bibnamefont {Mustafa}}, \bibinfo {author}
  {\bibfnamefont {Y.}~\bibnamefont {Nohara}}, \bibinfo {author} {\bibfnamefont
  {Y.}~\bibnamefont {Nomura}}, \bibinfo {author} {\bibfnamefont
  {L.}~\bibnamefont {Paulatto}}, \bibinfo {author} {\bibfnamefont
  {S.}~\bibnamefont {Ponc{\'{e}}}}, \bibinfo {author} {\bibfnamefont
  {T.}~\bibnamefont {Ponweiser}}, \bibinfo {author} {\bibfnamefont
  {J.}~\bibnamefont {Qiao}}, \bibinfo {author} {\bibfnamefont {F.}~\bibnamefont
  {Th\"{o}le}}, \bibinfo {author} {\bibfnamefont {S.~S.}\ \bibnamefont
  {Tsirkin}}, \bibinfo {author} {\bibfnamefont {M.}~\bibnamefont
  {Wierzbowska}}, \bibinfo {author} {\bibfnamefont {N.}~\bibnamefont
  {Marzari}}, \bibinfo {author} {\bibfnamefont {D.}~\bibnamefont {Vanderbilt}},
  \bibinfo {author} {\bibfnamefont {I.}~\bibnamefont {Souza}}, \bibinfo
  {author} {\bibfnamefont {A.~A.}\ \bibnamefont {Mostofi}},\ and\ \bibinfo
  {author} {\bibfnamefont {J.~R.}\ \bibnamefont {Yates}},\ }\bibfield  {title}
  {\bibinfo {title} {Wannier90 as a community code: new features and
  applications},\ }\href {https://doi.org/10.1088/1361-648x/ab51ff} {\bibfield
  {journal} {\bibinfo  {journal} {Journal of Physics: Condensed Matter}\
  }\textbf {\bibinfo {volume} {32}},\ \bibinfo {pages} {165902} (\bibinfo
  {year} {2020})}\BibitemShut {NoStop}%
\bibitem [{\citenamefont {Korotin}\ \emph {et~al.}(2015)\citenamefont
  {Korotin}, \citenamefont {Mazurenko}, \citenamefont {Anisimov},\ and\
  \citenamefont {Streltsov}}]{korotin2015prb}%
  \BibitemOpen
  \bibfield  {author} {\bibinfo {author} {\bibfnamefont {D.~M.}\ \bibnamefont
  {Korotin}}, \bibinfo {author} {\bibfnamefont {V.~V.}\ \bibnamefont
  {Mazurenko}}, \bibinfo {author} {\bibfnamefont {V.~I.}\ \bibnamefont
  {Anisimov}},\ and\ \bibinfo {author} {\bibfnamefont {S.~V.}\ \bibnamefont
  {Streltsov}},\ }\bibfield  {title} {\bibinfo {title} {Calculation of exchange
  constants of the {H}eisenberg model in plane-wave-based methods using the
  {G}reen's function approach},\ }\href
  {https://doi.org/10.1103/PhysRevB.91.224405} {\bibfield  {journal} {\bibinfo
  {journal} {Phys. Rev. B}\ }\textbf {\bibinfo {volume} {91}},\ \bibinfo
  {pages} {224405} (\bibinfo {year} {2015})}\BibitemShut {NoStop}%
\bibitem [{\citenamefont {Blanco-Rey}\ \emph {et~al.}(2019)\citenamefont
  {Blanco-Rey}, \citenamefont {Cerdá},\ and\ \citenamefont
  {Arnau}}]{blanco-rey2019njp}%
  \BibitemOpen
  \bibfield  {author} {\bibinfo {author} {\bibfnamefont {M.}~\bibnamefont
  {Blanco-Rey}}, \bibinfo {author} {\bibfnamefont {J.~I.}\ \bibnamefont
  {Cerdá}},\ and\ \bibinfo {author} {\bibfnamefont {A.}~\bibnamefont
  {Arnau}},\ }\bibfield  {title} {\bibinfo {title} {Validity of perturbative
  methods to treat the spin-orbit interaction: application to
  magnetocrystalline anisotropy},\ }\href
  {https://doi.org/10.1088/1367-2630/ab3060} {\bibfield  {journal} {\bibinfo
  {journal} {New Journal of Physics}\ }\textbf {\bibinfo {volume} {21}},\
  \bibinfo {pages} {073054} (\bibinfo {year} {2019})}\BibitemShut {NoStop}%
\bibitem [{\citenamefont {Rosenberg}\ \emph {et~al.}(2022)\citenamefont
  {Rosenberg}, \citenamefont {DeStefano}, \citenamefont {Guo}, \citenamefont
  {Oh}, \citenamefont {Hashimoto}, \citenamefont {Lu}, \citenamefont
  {Birgeneau}, \citenamefont {Lee}, \citenamefont {Ke}, \citenamefont {Yi},\
  and\ \citenamefont {Chu}}]{rosenberg2022prb}%
  \BibitemOpen
  \bibfield  {author} {\bibinfo {author} {\bibfnamefont {E.}~\bibnamefont
  {Rosenberg}}, \bibinfo {author} {\bibfnamefont {J.~M.}\ \bibnamefont
  {DeStefano}}, \bibinfo {author} {\bibfnamefont {Y.}~\bibnamefont {Guo}},
  \bibinfo {author} {\bibfnamefont {J.~S.}\ \bibnamefont {Oh}}, \bibinfo
  {author} {\bibfnamefont {M.}~\bibnamefont {Hashimoto}}, \bibinfo {author}
  {\bibfnamefont {D.}~\bibnamefont {Lu}}, \bibinfo {author} {\bibfnamefont
  {R.~J.}\ \bibnamefont {Birgeneau}}, \bibinfo {author} {\bibfnamefont
  {Y.}~\bibnamefont {Lee}}, \bibinfo {author} {\bibfnamefont {L.}~\bibnamefont
  {Ke}}, \bibinfo {author} {\bibfnamefont {M.}~\bibnamefont {Yi}},\ and\
  \bibinfo {author} {\bibfnamefont {J.-H.}\ \bibnamefont {Chu}},\ }\bibfield
  {title} {\bibinfo {title} {Uniaxial ferromagnetism in the kagome metal
  {TbV$_{6}$Sn$_6$}},\ }\href {https://doi.org/10.1103/PhysRevB.106.115139}
  {\bibfield  {journal} {\bibinfo  {journal} {Phys. Rev. B}\ }\textbf {\bibinfo
  {volume} {106}},\ \bibinfo {pages} {115139} (\bibinfo {year} {2022})},\
  \bibinfo {note} {(\textit{Editors' Suggestion})}\BibitemShut {NoStop}%
\bibitem [{\citenamefont {Lee}\ \emph {et~al.}(2023)\citenamefont {Lee},
  \citenamefont {Skomski}, \citenamefont {Wang}, \citenamefont {Orth},
  \citenamefont {Ren}, \citenamefont {Kang}, \citenamefont {Pathak},
  \citenamefont {Kutepov}, \citenamefont {Harmon}, \citenamefont {McQueeney},
  \citenamefont {Mazin},\ and\ \citenamefont {Ke}}]{lee2023prb}%
  \BibitemOpen
  \bibfield  {author} {\bibinfo {author} {\bibfnamefont {Y.}~\bibnamefont
  {Lee}}, \bibinfo {author} {\bibfnamefont {R.}~\bibnamefont {Skomski}},
  \bibinfo {author} {\bibfnamefont {X.}~\bibnamefont {Wang}}, \bibinfo {author}
  {\bibfnamefont {P.~P.}\ \bibnamefont {Orth}}, \bibinfo {author}
  {\bibfnamefont {Y.}~\bibnamefont {Ren}}, \bibinfo {author} {\bibfnamefont
  {B.}~\bibnamefont {Kang}}, \bibinfo {author} {\bibfnamefont {A.~K.}\
  \bibnamefont {Pathak}}, \bibinfo {author} {\bibfnamefont {A.}~\bibnamefont
  {Kutepov}}, \bibinfo {author} {\bibfnamefont {B.~N.}\ \bibnamefont {Harmon}},
  \bibinfo {author} {\bibfnamefont {R.~J.}\ \bibnamefont {McQueeney}}, \bibinfo
  {author} {\bibfnamefont {I.~I.}\ \bibnamefont {Mazin}},\ and\ \bibinfo
  {author} {\bibfnamefont {L.}~\bibnamefont {Ke}},\ }\bibfield  {title}
  {\bibinfo {title} {Interplay between magnetism and band topology in the
  kagome magnets {$R$Mn$_6$Sn$_6$}},\ }\href
  {https://doi.org/10.1103/PhysRevB.108.045132} {\bibfield  {journal} {\bibinfo
   {journal} {Phys. Rev. B}\ }\textbf {\bibinfo {volume} {108}},\ \bibinfo
  {pages} {045132} (\bibinfo {year} {2023})}\BibitemShut {NoStop}%
\bibitem [{\citenamefont {Timmons}\ \emph {et~al.}(2020)\citenamefont
  {Timmons}, \citenamefont {Teknowijoyo}, \citenamefont {Ko\'{n}czykowski},
  \citenamefont {Cavani}, \citenamefont {Tanatar}, \citenamefont {Ghimire},
  \citenamefont {Cho}, \citenamefont {Lee}, \citenamefont {Ke}, \citenamefont
  {Jo}, \citenamefont {Bud'ko}, \citenamefont {Canfield}, \citenamefont {Orth},
  \citenamefont {Scheurer},\ and\ \citenamefont {Prozorov}}]{timmons2020prr}%
  \BibitemOpen
  \bibfield  {author} {\bibinfo {author} {\bibfnamefont {E.~I.}\ \bibnamefont
  {Timmons}}, \bibinfo {author} {\bibfnamefont {S.}~\bibnamefont
  {Teknowijoyo}}, \bibinfo {author} {\bibfnamefont {M.}~\bibnamefont
  {Ko\'{n}czykowski}}, \bibinfo {author} {\bibfnamefont {O.}~\bibnamefont
  {Cavani}}, \bibinfo {author} {\bibfnamefont {M.~A.}\ \bibnamefont {Tanatar}},
  \bibinfo {author} {\bibfnamefont {S.}~\bibnamefont {Ghimire}}, \bibinfo
  {author} {\bibfnamefont {K.}~\bibnamefont {Cho}}, \bibinfo {author}
  {\bibfnamefont {Y.}~\bibnamefont {Lee}}, \bibinfo {author} {\bibfnamefont
  {L.}~\bibnamefont {Ke}}, \bibinfo {author} {\bibfnamefont {N.~H.}\
  \bibnamefont {Jo}}, \bibinfo {author} {\bibfnamefont {S.~L.}\ \bibnamefont
  {Bud'ko}}, \bibinfo {author} {\bibfnamefont {P.~C.}\ \bibnamefont
  {Canfield}}, \bibinfo {author} {\bibfnamefont {P.~P.}\ \bibnamefont {Orth}},
  \bibinfo {author} {\bibfnamefont {M.~S.}\ \bibnamefont {Scheurer}},\ and\
  \bibinfo {author} {\bibfnamefont {R.}~\bibnamefont {Prozorov}},\ }\bibfield
  {title} {\bibinfo {title} {{Electron irradiation effects on superconductivity
  in ${\mathrm{PdTe}}_{2}$: An application of a generalized Anderson
  theorem}},\ }\href {https://doi.org/10.1103/PhysRevResearch.2.023140}
  {\bibfield  {journal} {\bibinfo  {journal} {Phys. Rev. Research}\ }\textbf
  {\bibinfo {volume} {2}},\ \bibinfo {pages} {023140} (\bibinfo {year}
  {2020})}\BibitemShut {NoStop}%
\bibitem [{\citenamefont {Holstein}\ and\ \citenamefont
  {Primakoff}(1940)}]{holstein1940pr}%
  \BibitemOpen
  \bibfield  {author} {\bibinfo {author} {\bibfnamefont {T.}~\bibnamefont
  {Holstein}}\ and\ \bibinfo {author} {\bibfnamefont {H.}~\bibnamefont
  {Primakoff}},\ }\bibfield  {title} {\bibinfo {title} {Field dependence of the
  intrinsic domain magnetization of a ferromagnet},\ }\href
  {https://doi.org/10.1103/PhysRev.58.1098} {\bibfield  {journal} {\bibinfo
  {journal} {Phys. Rev.}\ }\textbf {\bibinfo {volume} {58}},\ \bibinfo {pages}
  {1098} (\bibinfo {year} {1940})}\BibitemShut {NoStop}%
\bibitem [{\citenamefont {Toth}\ and\ \citenamefont
  {Lake}(2015)}]{toth2015jpcm}%
  \BibitemOpen
  \bibfield  {author} {\bibinfo {author} {\bibfnamefont {S.}~\bibnamefont
  {Toth}}\ and\ \bibinfo {author} {\bibfnamefont {B.}~\bibnamefont {Lake}},\
  }\bibfield  {title} {\bibinfo {title} {Linear spin wave theory for single-q
  incommensurate magnetic structures},\ }\href
  {https://doi.org/10.1088/0953-8984/27/16/166002} {\bibfield  {journal}
  {\bibinfo  {journal} {Journal of Physics: Condensed Matter}\ }\textbf
  {\bibinfo {volume} {27}},\ \bibinfo {pages} {166002} (\bibinfo {year}
  {2015})}\BibitemShut {NoStop}%
\bibitem [{\citenamefont {Abernathy}\ \emph {et~al.}(2012)\citenamefont
  {Abernathy}, \citenamefont {Stone}, \citenamefont {Loguillo}, \citenamefont
  {Lucas}, \citenamefont {Delaire}, \citenamefont {Tang}, \citenamefont {Lin},\
  and\ \citenamefont {Fultz}}]{Abernathy2012}%
  \BibitemOpen
  \bibfield  {author} {\bibinfo {author} {\bibfnamefont {D.~L.}\ \bibnamefont
  {Abernathy}}, \bibinfo {author} {\bibfnamefont {M.~B.}\ \bibnamefont
  {Stone}}, \bibinfo {author} {\bibfnamefont {M.~J.}\ \bibnamefont {Loguillo}},
  \bibinfo {author} {\bibfnamefont {M.~S.}\ \bibnamefont {Lucas}}, \bibinfo
  {author} {\bibfnamefont {O.}~\bibnamefont {Delaire}}, \bibinfo {author}
  {\bibfnamefont {X.}~\bibnamefont {Tang}}, \bibinfo {author} {\bibfnamefont
  {J.~Y.~Y.}\ \bibnamefont {Lin}},\ and\ \bibinfo {author} {\bibfnamefont
  {B.}~\bibnamefont {Fultz}},\ }\bibfield  {title} {\bibinfo {title} {{Design
  and operation of the wide angular-range chopper spectrometer ARCS at the
  Spallation Neutron Source}},\ }\bibfield  {journal} {\bibinfo  {journal}
  {Review of Scientific Instruments}\ }\textbf {\bibinfo {volume} {83}},\ \href
  {https://doi.org/10.1063/1.3680104} {10.1063/1.3680104} (\bibinfo {year}
  {2012}),\ \bibinfo {note} {015114}\BibitemShut {NoStop}%
\bibitem [{\citenamefont {Li}(2023)}]{pyLiSW}%
  \BibitemOpen
  \bibfield  {author} {\bibinfo {author} {\bibfnamefont {B.}~\bibnamefont
  {Li}},\ }\href {https://doi.org/10.5281/zenodo.8157455} {\bibinfo {title}
  {bingli621/pylisw: pylisw v1.0}} (\bibinfo {year} {2023})\BibitemShut
  {NoStop}%
\bibitem [{\citenamefont {Liu}\ \emph {et~al.}(2017)\citenamefont {Liu},
  \citenamefont {Hu}, \citenamefont {Zhang}, \citenamefont {Graf},
  \citenamefont {Cao}, \citenamefont {Radmanesh}, \citenamefont {Adams},
  \citenamefont {Zhu}, \citenamefont {Cheng}, \citenamefont {Liu},
  \citenamefont {Phelan}, \citenamefont {Wei}, \citenamefont {Jaime},
  \citenamefont {Balakirev}, \citenamefont {Tennant}, \citenamefont {DiTusa},
  \citenamefont {Chiorescu}, \citenamefont {Spinu},\ and\ \citenamefont
  {Mao}}]{liu2017nm}%
  \BibitemOpen
  \bibfield  {author} {\bibinfo {author} {\bibfnamefont {J.~Y.}\ \bibnamefont
  {Liu}}, \bibinfo {author} {\bibfnamefont {J.}~\bibnamefont {Hu}}, \bibinfo
  {author} {\bibfnamefont {Q.}~\bibnamefont {Zhang}}, \bibinfo {author}
  {\bibfnamefont {D.}~\bibnamefont {Graf}}, \bibinfo {author} {\bibfnamefont
  {H.~B.}\ \bibnamefont {Cao}}, \bibinfo {author} {\bibfnamefont {S.~M.~A.}\
  \bibnamefont {Radmanesh}}, \bibinfo {author} {\bibfnamefont {D.~J.}\
  \bibnamefont {Adams}}, \bibinfo {author} {\bibfnamefont {Y.~L.}\ \bibnamefont
  {Zhu}}, \bibinfo {author} {\bibfnamefont {G.~F.}\ \bibnamefont {Cheng}},
  \bibinfo {author} {\bibfnamefont {X.}~\bibnamefont {Liu}}, \bibinfo {author}
  {\bibfnamefont {W.~A.}\ \bibnamefont {Phelan}}, \bibinfo {author}
  {\bibfnamefont {J.}~\bibnamefont {Wei}}, \bibinfo {author} {\bibfnamefont
  {M.}~\bibnamefont {Jaime}}, \bibinfo {author} {\bibfnamefont
  {F.}~\bibnamefont {Balakirev}}, \bibinfo {author} {\bibfnamefont {D.~A.}\
  \bibnamefont {Tennant}}, \bibinfo {author} {\bibfnamefont {J.~F.}\
  \bibnamefont {DiTusa}}, \bibinfo {author} {\bibfnamefont {I.}~\bibnamefont
  {Chiorescu}}, \bibinfo {author} {\bibfnamefont {L.}~\bibnamefont {Spinu}},\
  and\ \bibinfo {author} {\bibfnamefont {Z.~Q.}\ \bibnamefont {Mao}},\
  }\bibfield  {title} {\bibinfo {title} {A magnetic topological semimetal
  {Sr}$_{1-y}${Mn}$_{1-z}${Sb}$_2$ (y, z $<$ 0.1)},\ }\href
  {https://doi.org/10.1038/nmat4953} {\bibfield  {journal} {\bibinfo  {journal}
  {Nature Materials}\ }\textbf {\bibinfo {volume} {16}},\ \bibinfo {pages}
  {905} (\bibinfo {year} {2017})}\BibitemShut {NoStop}%
\bibitem [{\citenamefont {Islam}\ \emph {et~al.}(2020)\citenamefont {Islam},
  \citenamefont {Choudhary}, \citenamefont {Liu}, \citenamefont {Ueland},
  \citenamefont {Paudyal}, \citenamefont {Heitmann}, \citenamefont
  {McQueeney},\ and\ \citenamefont {Vaknin}}]{islam2020prb}%
  \BibitemOpen
  \bibfield  {author} {\bibinfo {author} {\bibfnamefont {F.}~\bibnamefont
  {Islam}}, \bibinfo {author} {\bibfnamefont {R.}~\bibnamefont {Choudhary}},
  \bibinfo {author} {\bibfnamefont {Y.}~\bibnamefont {Liu}}, \bibinfo {author}
  {\bibfnamefont {B.~G.}\ \bibnamefont {Ueland}}, \bibinfo {author}
  {\bibfnamefont {D.}~\bibnamefont {Paudyal}}, \bibinfo {author} {\bibfnamefont
  {T.}~\bibnamefont {Heitmann}}, \bibinfo {author} {\bibfnamefont {R.~J.}\
  \bibnamefont {McQueeney}},\ and\ \bibinfo {author} {\bibfnamefont
  {D.}~\bibnamefont {Vaknin}},\ }\bibfield  {title} {\bibinfo {title}
  {Controlling magnetic order, magnetic anisotropy, and band topology in the
  semimetals{Sr(Mn}$_{0.9}${Cu}$_{0.1}$){Sb}$_{2}$ and
  {Sr(Mn}$_{0.9}${Zn}$_{0.1}$){Sb}$_{2}$},\ }\href
  {https://link.aps.org/doi/10.1103/PhysRevB.102.085130} {\bibfield  {journal}
  {\bibinfo  {journal} {Phys. Rev. B}\ }\textbf {\bibinfo {volume} {102}},\
  \bibinfo {pages} {085130} (\bibinfo {year} {2020})}\BibitemShut {NoStop}%
\bibitem [{\citenamefont {Liu}\ \emph {et~al.}(2019)\citenamefont {Liu},
  \citenamefont {Islam}, \citenamefont {Dennis}, \citenamefont {Tian},
  \citenamefont {Ueland}, \citenamefont {McQueeney},\ and\ \citenamefont
  {Vaknin}}]{Liu2019prbhd}%
  \BibitemOpen
  \bibfield  {author} {\bibinfo {author} {\bibfnamefont {Y.}~\bibnamefont
  {Liu}}, \bibinfo {author} {\bibfnamefont {F.}~\bibnamefont {Islam}}, \bibinfo
  {author} {\bibfnamefont {K.~W.}\ \bibnamefont {Dennis}}, \bibinfo {author}
  {\bibfnamefont {W.}~\bibnamefont {Tian}}, \bibinfo {author} {\bibfnamefont
  {B.~G.}\ \bibnamefont {Ueland}}, \bibinfo {author} {\bibfnamefont {R.~J.}\
  \bibnamefont {McQueeney}},\ and\ \bibinfo {author} {\bibfnamefont
  {D.}~\bibnamefont {Vaknin}},\ }\bibfield  {title} {\bibinfo {title} {Hole
  doping and antiferromagnetic correlations above the {N}\'{e}el temperature of
  the topological semimetal ({Sr}$_{1-x}${K}$_{x}$){MnSb}$_{2}$},\ }\href
  {https://doi.org/10.1103/PhysRevB.100.014437} {\bibfield  {journal} {\bibinfo
   {journal} {Phys. Rev. B}\ }\textbf {\bibinfo {volume} {100}},\ \bibinfo
  {pages} {014437} (\bibinfo {year} {2019})}\BibitemShut {NoStop}%
\bibitem [{sup()}]{suppl_srmnsb2}%
  \BibitemOpen
  \href@noop {} {}\bibinfo {note} {See Supplemental Material [url] for
  symmetrized exchange interactions, theoretical calculation of spin spectra
  and INS data treatment and modeling.}\BibitemShut {Stop}%
\end{thebibliography}

\begin{thebibliography}{11}
%\bibitem{S_RefA} A. Someone, C. Someone, D. Someone, Phys. Rev. Lett. {\bf 11}, 1111 (1911).
\end{thebibliography}
%\bigskip


\clearpage

\title{\large Magnetic interactions and excitations in SrMnSb$_2$\\Supplemental Materials}

\author{Zhenhua Ning}
\affiliation{Ames National Laboratory, U.S. Department of Energy, Ames, Iowa 50011}
\author{Bing Li}
\affiliation{Ames National Laboratory, U.S. Department of Energy, Ames, Iowa 50011}
%% \author{V. V. MKhitaryan}
\author{Arnab Banerjee} % instrumental scientist at that time
\affiliation{Department of Physics and Astronomy, Purdue University, West Lafayette, IN 47906}
\author{Victor Fanelli} %  ARCS scientist associate at that time
\author{Doug Abernathy} % instrumental scientist 
\affiliation{Neutron Scattering Division, Oak Ridge National Laboratory, Oak Ridge, TN, 37831}
\author{Yong Liu} %sample grower
\author{Benjamin G Ueland}
\affiliation{Ames National Laboratory, U.S. Department of Energy, Ames, Iowa 50011}
\author{Robert J. McQueeney}
\affiliation{Ames National Laboratory, U.S. Department of Energy, Ames, Iowa 50011}
\author{Liqin Ke}
\affiliation{Ames National Laboratory, U.S. Department of Energy, Ames, Iowa 50011}
%%%%%%%%%% Merge with supplemental materials %%%%%%%%%%
\maketitle
\widetext
\clearpage
%\begin{center}
%\textbf{\large Supplemental Materials: Magnetic interactions and excitations in SrMnSb$_2$}
%\end{center}
%%%%%%%%%% Merge with supplemental materials %%%%%%%%%%
%%%%%%%%%% Prefix a "S" to all equations, figures, tables and reset the counter %%%%%%%%%%
\setcounter{equation}{0}
\setcounter{figure}{0}
\setcounter{table}{0}
\setcounter{page}{1}
\makeatletter
\renewcommand{\theequation}{S\arabic{equation}}
\renewcommand{\thefigure}{S\arabic{figure}}
\renewcommand{\bibnumfmt}[1]{[S#1]}
\renewcommand{\citenumfont}[1]{S#1}
%%%%%%%%%% Prefix a "S" to all equations, figures, tables and reset the counter %%%%%%%%%%

\section{Calculated spin wave spectra and symmetrized $J_1$}

\begin{figure}[ht]
\centering
\begin{tabular}{c}
\includegraphics[width=0.69\linewidth,clip]{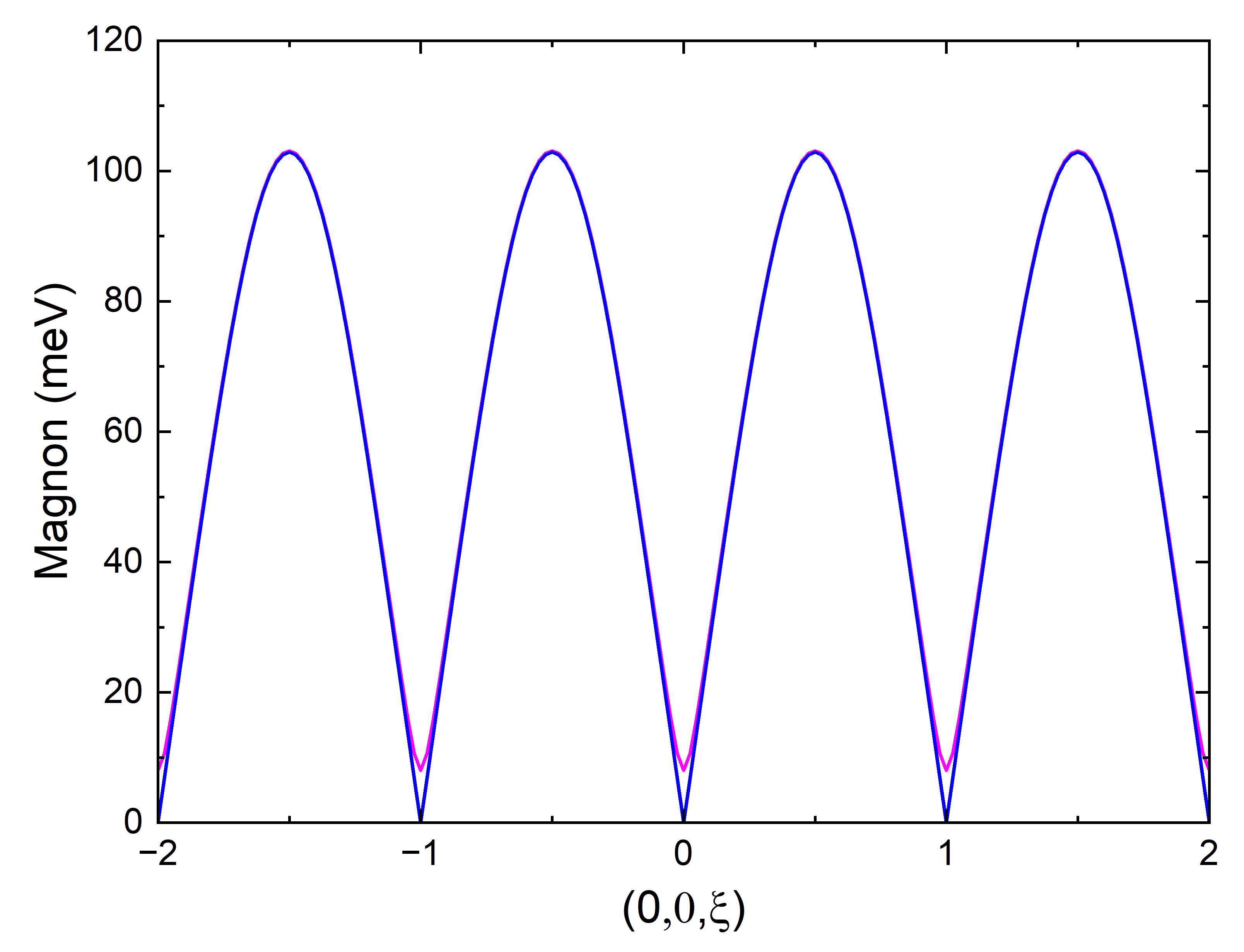} \\
\includegraphics[width=0.69\linewidth,clip]{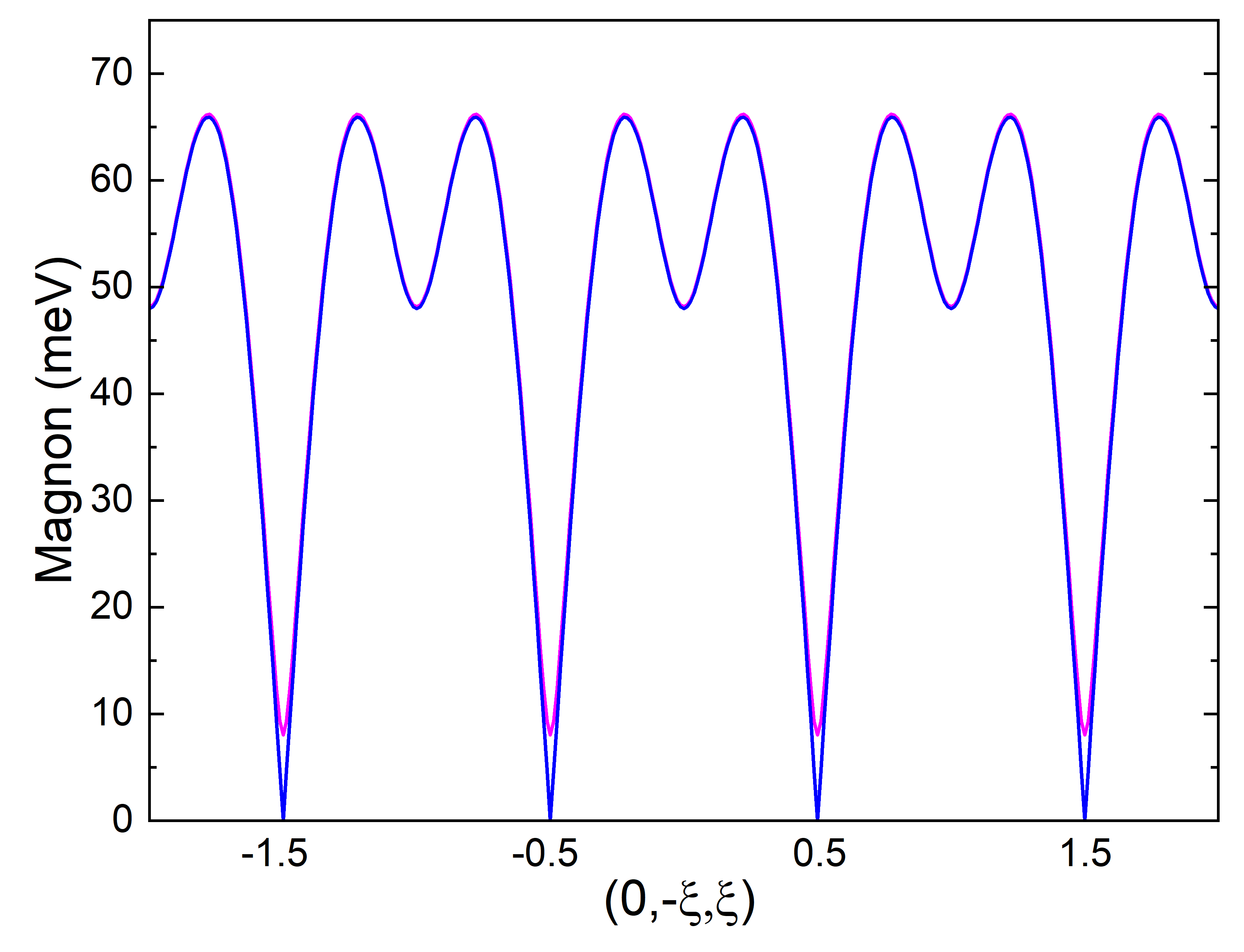} \\   
\end{tabular}%
\caption{ Spinwave spectra of $\srmnsb$. Top panel is the spin wave dispersion along $(0,0,\xi)$ and bottom panel along $(0,-\xi,\xi)$.}
\label{fig:spinwave}
\end{figure}

With exchange parameters obtained by linear response theory, the spectra of spin wave is calculated using $J_1-J_2-J_a$ model given by
\begin{eqnarray}\label{J1J2JaHam}
  H=\sum_{<i,j>}^{NN} J_{1} \hat{\bold{S}}_{i}\cdot \hat{\bold{S}}_{j}+\sum_{<i,j>}^{NNN} J_{2} \hat{\bold{S}}_{i}\cdot \hat{\bold{S}}_{j}+\sum_{<i,j>} J_{a} \bold{S}_{i}\cdot \bold{S}_{j}+ \sum_{i} D(S_z)^2.\,,
\end{eqnarray}

\begin{table}[bhtp]
\caption{Real-space pairwise exchange parameters $\jij$ of $\srmnsb$ for the Heisenberg Hamiltonian for the star of neighbors, marked by primitive translation vectors $\bold R$. 
The first row is calculated with the symmetrized Hamiltonian while the second one is done with the non-symmetrized Hamiltonian. 
To show symmetry clearly, numbers assume up to the fifth decimal place.}
\label{tbl:j12sym}%
\bgroup
\def\arraystretch{1.1}% 
  \begin{tabular*}{\linewidth}{l @{\extracolsep{\fill}} rrrrr}
  \hline\hline
  \\[-1em]
     $\bold R$                            &   & [0,0,0]  & [0,-1,0] &  [0,0,-1]  &  [0,-1,-1]   \\   \\[-1.1em]   \hline    \\[-1.1em]
  $J_1^{sym}$ (\si{meV})          &   & 11.23149 &   11.23149   &  11.23149 &  11.23149  \\  
  $J_1$ (\si{meV})              &   & 11.34235 &   11.32793   &  11.33891 &  11.34101  \\
  \\[-1.1em]
  \hline\hline
\end{tabular*}
\egroup
\end{table}

Table \ref{tbl:j12sym} presents the intra-sublattice interactions $J_1$ between nearest neighbors (NN) with a bond length of \SI{3.14}{\angstrom}.
It is evident that the exchange parameters calculated using the symmetrized Hamiltonian are identical for the star of neighbors, whereas those obtained using the non-symmetrized Hamiltonian differ at the second decimal place.
Despite the lack of symmetry, both sets of values agree within $0.6\%$. 
The intra-sublattice interactions between the next-nearest neighbors (NNN) exhibit similar symmetric behavior, with an average value of \SI{7.12}{meV}. This value is less than half of $J_1$, indicating frustration in NNN interactions.

%\section{DOS of Mn site and Orbital resolved exchange coupling}
%\label{appen:DOS_Mn_J1w_J2w}
% The projected DOS for one Mn site and the orbital resolved exchange coupling as a function of energy are plotted in Figure \ref{fig:DOS_Mn_J1w_J2w}.
%\begin{figure*}[ht]
% \centering
% \includegraphics[width=.60\linewidth,clip]{SrMnSb2_pdos_bca.png}
% \includegraphics[width=.60\linewidth,clip]{SrMnSb2_j1w_bca.png}
% \includegraphics[width=.60\linewidth,clip]{SrMnSb2_j2w_bca.png}
% \caption{ Panel (a) illustrates the projected DOS for a single Mn site. In contrast, Panels (b) and (c) depict the orbital-resolved exchange coupling as a function of energy.}
% \label{fig:DOS_Mn_J1w_J2w}
%\end{figure*}

\section{Details of INS data treatment and modeling}
\label{appen:INS_details}

\begin{figure*}[ht]
  \centering
  \includegraphics[width=.90\linewidth,clip]{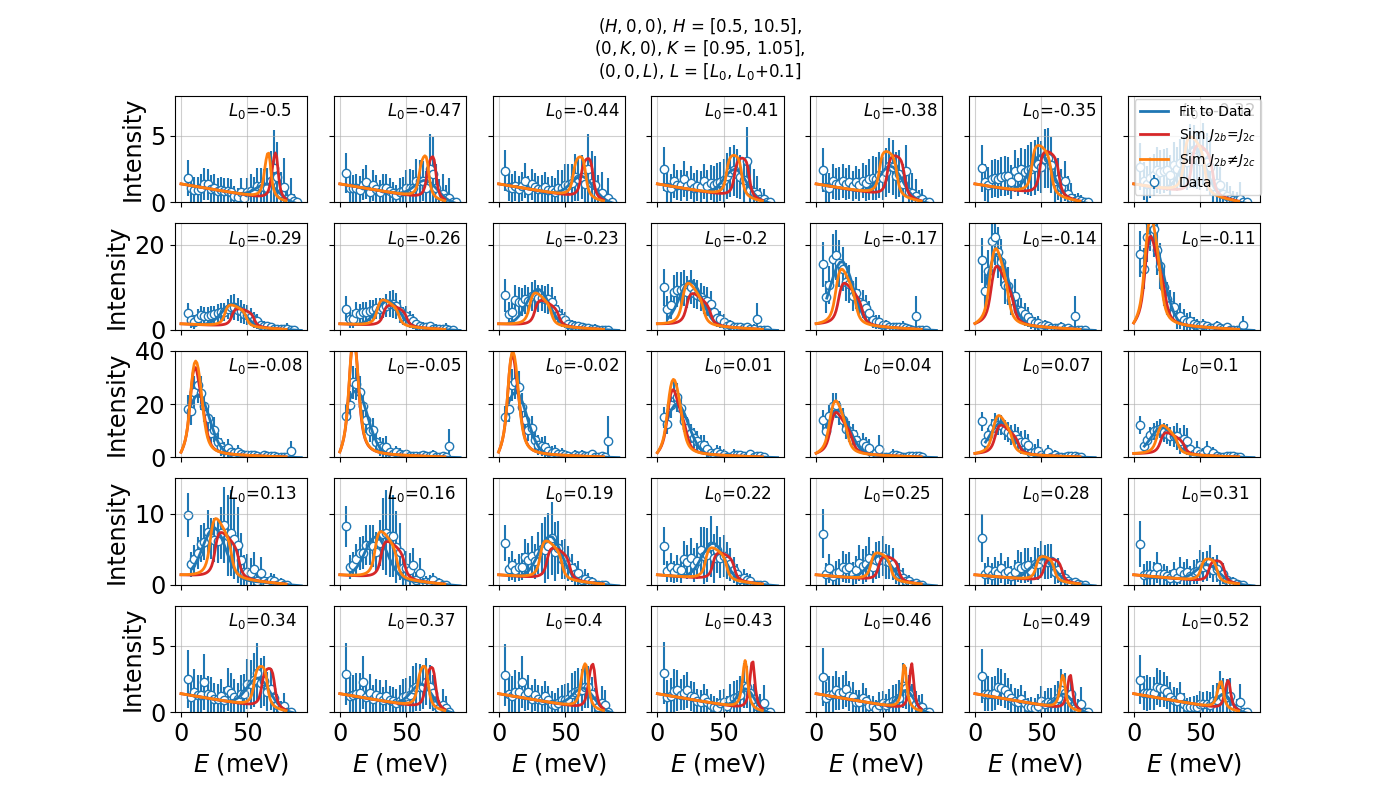}
  \caption{ Cuts of inelastic neutron scattering along (0,1,L) made using the integration ranges indicated. Lines show fits using Gaussian functions and simulations using the two model Hamiltonians discussed in the text.}
  \label{fig:INS_fit_001}
\end{figure*}

\begin{figure*}[ht]
  \centering
  \includegraphics[width=.90\linewidth,clip]{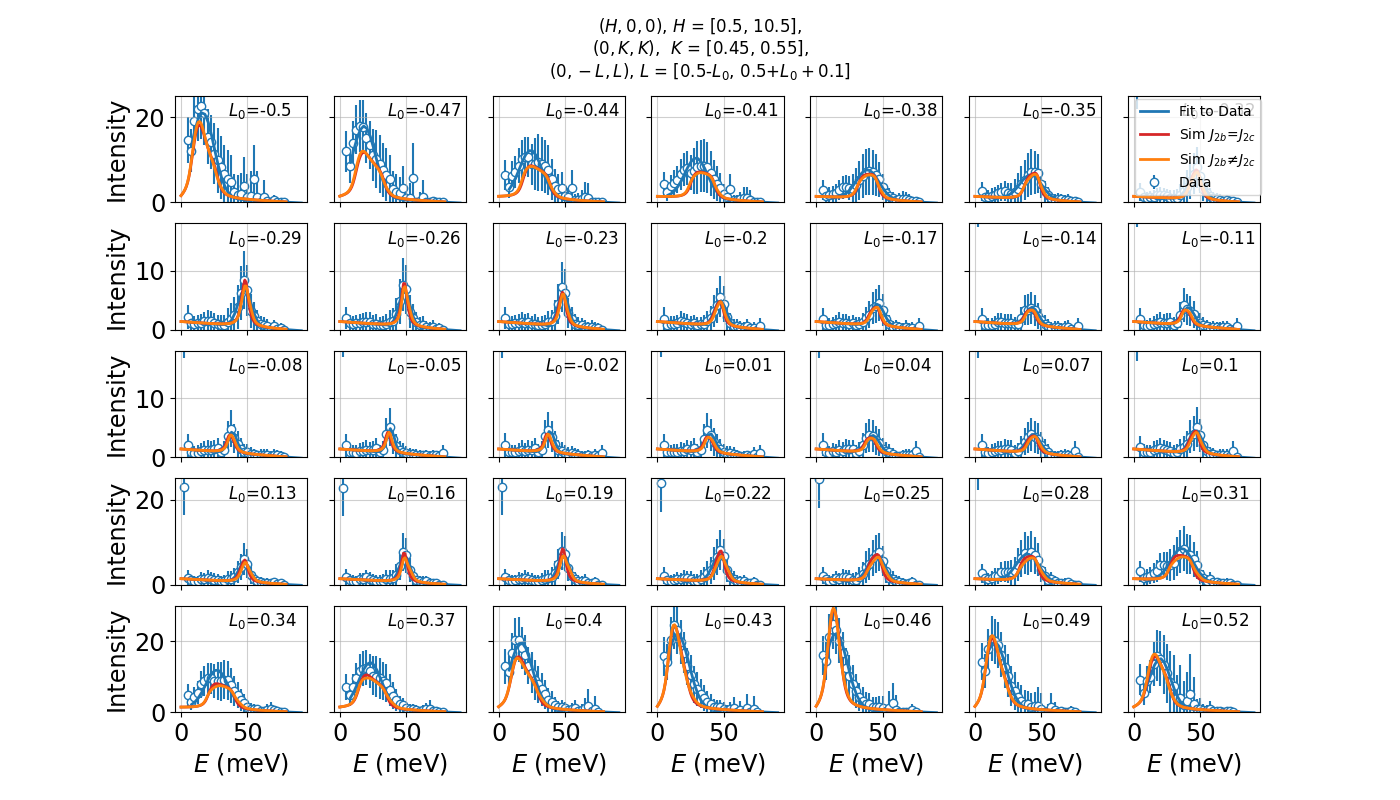}
  \caption{Cuts of inelastic neutron scattering along (0,1/2-K,1/2+K)   made using the integration ranges indicated. Lines show fits using Gaussian functions and simulations using the two model Hamiltonians discussed in the text.}
  \label{fig:INS_fit_011}
\end{figure*}

\begin{figure*}[ht]
  \centering
  \includegraphics[width=.90\linewidth,clip]{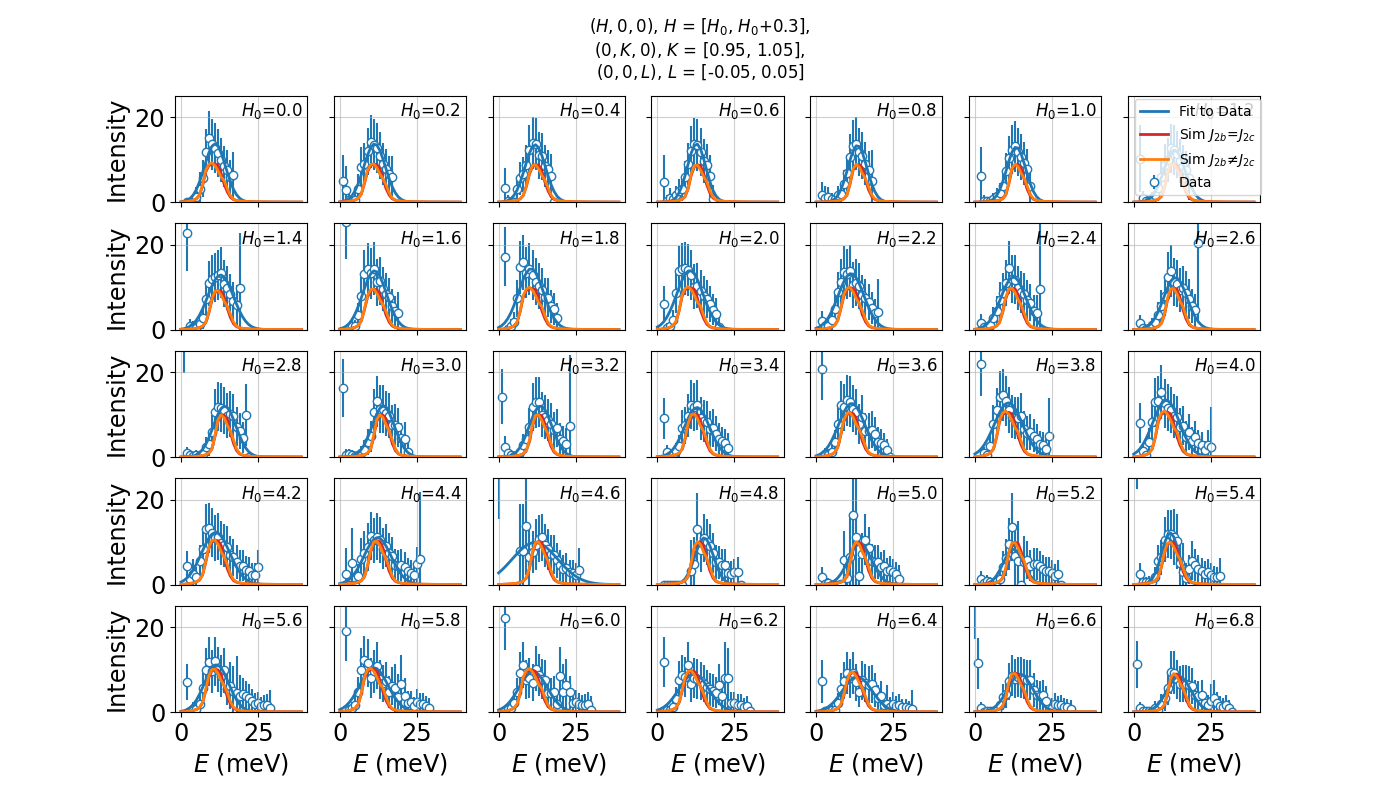}
  \caption{Cuts of inelastic neutron scattering along (H,1,0)   made using the integration ranges indicated. Lines show fits using Gaussian functions and simulations using the two model Hamiltonians discussed in the text.}
  \label{fig:INS_fit_100}
\end{figure*}

We focus on the three high-symmetry directions $(1,0,L)$, $(0,0.5-K, 0.5+K)$ and $(H,1,0)$, and plotted the spectra as functions of energy transfer $E$ as shown in  Figs.~\ref{fig:INS_fit_001}, \ref{fig:INS_fit_011}, and \ref{fig:INS_fit_100}, respectively. The bin ranges are shown in the title of each plot. The line cuts were fit to a Gaussian, shown by the blue curves, to extract the peak centers as shown in Fig.9 in the main text. %~\ref{fig:INS_disp}. 
In Figs.~\ref{fig:INS_fit_001} and \ref{fig:INS_fit_011}, a $\mathbf{Q}$-independent linear background was also introduced to account for the weak phonon signals from the aluminum sample holder.

To model the spin wave data, we chose the chemical unit cell with 4 Mn ions, and introduced in-plane interactions $J_1$, $J_2$, inter-layer interaction $J_{a}$, and an uniaxial single-ion anisotropy $D$. The magnetic Hamiltonian is given by Eq.~\eqref{J1J2JaHam}.

We implemented the linear spin wave theory to calculate the spin wave dispersion and INS spectra. We used the peak shape of the under-damped harmonic oscillator to model the INS spectra at each $\mathbf{Q}$ point, where the damping parameters are chosen to be FWHM of the energy-dependent instrumental resolution of ARCS. We perform the same $\mathbf{Q}$-binning as in the data, and fit the energy line cuts to Gaussians to extract the peak centers. We then perform a second round of fitting to find the magnetic interactions that best match the extracted peaks centers in the data and the model.

Figs.~\ref{fig:INS_fit_001}, \ref{fig:INS_fit_011}, and \ref{fig:INS_fit_100} show the data compared to two models. Red curves were calculated from the fitting of peak centers while enforcing $J_{2b}=J_{2c}$, orange curves are from the fitting that keeps $J_{2b}$ and $J_{2c}$ as free parameters.

\end{document}